\documentclass[11pt]{article}
\usepackage{citesort,bibmods}
\usepackage{url}
\usepackage{amsmath}
\usepackage{amscd}
\usepackage{amssymb,latexsym}
\usepackage{float}
\usepackage{deleq}

\usepackage[mathscr]{eucal}
\usepackage{a4}

\DeclareFontFamily{OT1}{rsfs}{}
\DeclareFontShape{OT1}{rsfs}{m}{n}{ <-7> rsfs5 <7-10> rsfs7 <10-> rsfs10}{}
\DeclareMathAlphabet{\mycal}{OT1}{rsfs}{m}{n}
\newcommand{\nn}{\nonumber}
\newcommand{\ncO}{{\cal O}}

\newcommand{\mcU}{{\mycal U}}
\newcommand{\cOp}{\widehat{\ncO}_{\hat p}}
\newcommand{\hOp}{\cOp}

\def\th{{e}}
\def\tH{{H}}
\def\tV{\tilde{V}}
\def\tK{{K}}

\newcommand{\oeps}{o(r^{-2\alpha})}
\newcommand{\oepsb}{o(r^{\beta-2\alpha})}
\newcommand{\myvareps}{}
\newcommand{\newa}{c}

\newcommand{\mnoteprzelicz}[1]{}
\renewcommand{\lrcorner}{\rfloor} 

\newcounter{mnotecount}[section]

\newcommand{\mnote}[1]{}

\newcommand{\FS}       
                  {F}

\newcommand{\HS} 
       {H_{\mbox{\scriptsize volume}}}

\newcommand{\ren}{\mbox{\scriptsize normalised}}
\newcommand{\HSren} 
       {{\HS^{\ren}}}

\newcommand{\hypext}{{\hyp_{\ext}}}

\newcommand{\Lagrangian}{Lagrangian } 

\newcommand{\reword}{\mnote{rewordings}}
\newcommand{\ptc}[1]{\mnote{\textbf{PTC:} #1}}

  {\catcode `\@=11 \global\let\AddToReset=\@addtoreset}
\AddToReset{equation}{section}
\renewcommand{\theequation}{\thesection.\arabic{equation}}

\newcommand{\lie}{{\pounds}} 
\def\varOmega{{\mathit{\Omega}}}

\newcommand{\rdS}{\rd S} 

\newcommand{\cU}{{\mycal U}}
\newcommand{\cP}{{\mycal P}}

\newtheorem{Theorem} {Theorem} [section] \newtheorem{Corollary}
[Theorem] {Corollary} \newtheorem{Lemma} [Theorem] {Lemma}
\newtheorem{Proposition} [Theorem] {Proposition}

\newcommand{\eq}[1]{(\ref{#1})}

      \newcommand{\be}{\begin{equation}}
      \newcommand{\ee}{\end{equation}}
\newcommand{\eea}{\end{eqnarray}}
\newcommand{\bea}{\begin{eqnarray}}

      \newfont{\msa}{msam10 scaled\magstep1}

\def\Reals{{\mathbb R}}
\def\T{{\mathbb T}}
 \def\Naturals{{\mathbb N}}

\newcommand{\cLX}{{\lie_X}}
\newcommand{\caD}{{{\mycal D}}}
\newcommand{\rd}{\,{ d}} 
\newcommand{\N}{\Naturals}
\newcommand{\R}{\Reals} 

\newcommand{\przecinekJped}{}

\newcommand{\hyp}{{\mycal S}}

\newcommand{\Bgamma}{{B}} 
\newcommand{\bmetric}{{b}} 

\newcommand{\Kp}{{\mathfrak g}} 
\newcommand{\KA}{p} 

\newcommand{\gthreeup}{\,{}^3g} 
\newcommand{\cO}{{\mathscr O}}

\newcommand{\ext}{{\mbox{\rm\scriptsize ext}}}

\newcommand{\bel}[1]{\be\label{#1}}
\newcommand{\ourU}{\mathbb U}

\newcommand{\hzome}{\widehat{\mathring{\omega}}}

\newcommand{\hhyp}{\,\,\widehat{\!\!\hyp}\,} 

\newcommand{\hM}{\,\widehat{\!M}} 

\newcommand{\pgp}{P_{\hM}\gamma_p}
\newcommand{\pgq}{P_{\hM}\gamma_q}

\newcommand{\zR} {\zRs} 
\newcommand{\zRs} {\mathring{R}} 
\newcommand{\zRm} {\mathring{R}} 
\newcommand{\zo} {\mathring{\omega}} 
\newcommand{\hzo} {\widehat{\zo}{}} 
\newcommand{\znabla} {\mathring{\nabla}} 
\newcommand{\sRz} {\mathring{\sigma}_R} 
\newcommand{\sR} {\sigma_R} 

\newcommand{\frg}{{\mathfrak g}} 
\newcommand{\fb}{{\mathfrak b}} 

\newcommand{\zn} {\mathring{\nabla}} 
\newcommand{\mn} {M} 
\newcommand{\cM}{\mycal M} 
\newcommand{\cK}{\mycal K} 
\newcommand{\hcK} {\,\,\widehat{\!\!\cK}} 
\newcommand{\cKS}{{\mycal K}_{\hyp^{\perp}}} 
\newcommand{\cKt}{{\mycal K}_{\hyp^{\parallel}}} 
\newcommand{\Eq}[1]{Equation~\eq{#1}}
\newcommand{\Eqsone}[1]{Equations~\eq{#1}}
\newcommand{\Eqs}[2]{Equations~\eq{#1}-\eq{#2}}
\newcommand{\hr}{{\hat r}}
\newcommand{\hbeta}{{\hat \beta}}
\newcommand{\hx}{{\hat x}}
\newcommand{\hphi}{{\hat \phi}}
\newcommand{\hth}{{\hat \theta}}
\newcommand{\htheta}{{\hth}}
\newcommand{\home}{{\hat \omega}}
\newcommand{\hh}{{\hat h}}
\newcommand{\hf}{{\hat f}}
\newcommand{\hb}{{\hat b}}
\newcommand{\hv}{{\hat v}}
\newcommand{\he}{{\hat e}}
\newcommand{\proof}{\noindent {\sl Proof:} }
\newcommand{\remark}{\noindent {\bf Remark:} }
\newcommand{\remarks}{\noindent {\bf Remarks:} }
\newcommand{\qed}{\hfill $\Box$\\ \medskip}
\newcommand{\ds}{\; ds}
\begin{document}\newcommand{\Sext}{\hyp_{{\text{\scriptsize\rm ext}}}}
\newcommand{\fg}{{}^4{g}}
\newcommand{\ac}{\zeta}

\title{The mass of spacelike hypersurfaces in asymptotically anti-de
  Sitter space-times}

\author{Piotr T.\ Chru\'sciel\thanks{ Supported in part by the Polish
    Research Council grant KBN 2 P03B 073 15, and by the A. von Humboldt foundation. E--mail:
    \texttt{chrusciel@univ-tours.fr}} \\ Gabriel Nagy\thanks{
    Current address: Max-Planck-Institut f\"ur Gravitationsphysik,
                    Albert-Einstein-Institut,
                    Am M\"uhlenberg 1,
                    D-14476 Golm,
                    Germany. Supported by a grant from R\'egion Centre. E--mail:
    \texttt{nagy@aei-potsdam.mpg.de}} \\
  D\'epartement de
  Math\'ematiques \\ Facult\'e des Sciences\\
  Parc de Grandmont \\F-37200 Tours, France}

\maketitle

\begin{abstract}
  We give a Hamiltonian definition of mass for spacelike hypersurfaces
  in space-times with metrics which are asymptotic to the anti-de
  Sitter one, or to a class of generalizations thereof. We show that
  our definition provides a geometric invariant for a spacelike
  hypersurface embedded in a space-time.
\end{abstract}
\tableofcontents

\newcommand{\mSigma}{\cal S}
\section{Introduction}
\label{Si} Let $\hyp$ be an $n$-dimensional spacelike hypersurface
in a $n+1$-dimensional Lorentzian space-time $(\cM,g)$, $n\ge 2$.
Suppose that $\cM$ contains an open set $\cU$ with a global time
coordinate $t$ (with range not necessarily equal to $\R$), as well
as a global ``radial'' coordinate $r\in[R,\infty)$, leading to
local coordinate systems $(t,r,v^A)$, with $(v^A)$ --- local
coordinates on some compact $n-1$ dimensional manifold $\mn$. We
further require that $\hyp\cap \cU=\{t=0\}$. Assume that the
metric $g$ approaches (as $r$ tends to infinity, in a sense which
is made precise in Section~\ref{S2} below) a background metric $b$
of the form
\begin{equation}
  \label{eq:a1}
  b= -{\textstyle{\left({r^2\over\ell^2}+k\right)}} dt^2 + {1\over{r^2\over\ell^2}+k } dr^2
  +r^2 h\;,
\end{equation}
where $h$ is an $r$-independent Riemannian metric on $\mn$, while
$k$ and $\ell$ are constants\footnote{\label{footWoolgar}A warped
product form of the metric, with the factor in front of $h$ not
being constant, together with the Einstein equations~\eq{Ee},
force $g_{rr}$ and $g_{tt}$ to have the form \eq{eq:a1} in an
appropriate coordinate system \cite{CadeauWoolgar}, with $k$ being
a function of $r$ which approaches a constant as $r$ tends to
infinity.  Further $h$ itself has to satisfy the Einstein
equation~\eq{Ee} with $\Lambda$ replaced by an appropriate
constant. Some metrics slightly more general than \eq{eq:a1} will
be considered in the body of the paper. }.
 Suppose further that $g$ satisfies the vacuum Einstein equations with
 a cosmological constant
\begin{equation}
  \label{Ee}
  R_{\mu\nu}-\frac{g^{\alpha \beta} R_{\alpha \beta}}{2} g_{\mu\nu} =
-\Lambda g_{\mu\nu} \;, \qquad\Lambda = -\textstyle{n(n-1) \over
2\ell^2}\;,
\end{equation}
similarly for $b$. (The existence of a large family of such $g$'s
follows from the work in \cite{Friedrich:adS,Kannar:adS}.) A
Hamiltonian analysis (following \cite{ChAIHP}, and discussed in
some more detail in Appendix \ref{AHam}; see also \cite[Section
5]{ChruscielSimon}) leads to the following expression for the
Hamiltonian associated to the flow of a vector field $X$, assumed
to be a Killing vector field of the background $b$:\footnote{The
integral over $\partial \hyp$ should be understood by a limiting
process, as the limit as $R$ tends to infinity of integrals over
the sets $t=0$, $r=R$. $d S_{\alpha\beta}$ is defined as
  $\frac{\partial}{\partial x^\alpha}\lrcorner
  \frac{\partial}{\partial x^\beta}\lrcorner \rd x^0 \wedge\cdots
  \wedge\rd x^{n} $, with $\lrcorner$ denoting contraction; $g$ stands
  for the space-time metric  unless explicitly indicated
  otherwise. Further, a semicolon denotes covariant differentiation
  \emph{with respect to the background metric $b$}.}
\begin{eqnarray}
  m(\hyp,g,b, X)&= &\frac 12 \int_{\partial\hyp}
 \ourU^{\alpha\beta}dS_{\alpha\beta}\;,
\label{toto}
\end{eqnarray}\begin{eqnarray}
  \ourU^{\nu\lambda}&= &
{\ourU^{\nu\lambda}}_{\beta}X^\beta + \frac 1{8\pi}
\left(\sqrt{|\det g_{\rho\sigma}|}~g^{\alpha[\nu}-\sqrt{|\det
b_{\rho\sigma}|}~ b^{\alpha[\nu}\right) {X^{\lambda]}}_{;\alpha} \
,\label{Fsup2new}
\\ {\ourU^{\nu\lambda}}_\beta &= & \displaystyle{\frac{2|\det
  \bmetric_{\mu\nu}|}{ 16\pi\sqrt{|\det g_{\rho\sigma}|}}}
g_{\beta\gamma}(e^2 g^{\gamma[\nu}g^{\lambda]\kappa})_{;\kappa}
\;,\label{Freud2.0} 
\\
  \label{mas2}
e&=& 
{\sqrt{|\det g_{\rho\sigma}|}}/{\sqrt{|\det\bmetric_{\mu\nu}|}}\;
.
\end{eqnarray}
(The question of convergence of the right-hand-side of \eq{toto}
is considered in Section~\ref{S2} below.) The hypersurface $\hyp$
singles out a set of Killing vectors $X$ for the metric $b$ which
are normal to $\hyp$,
\begin{equation}  \label{KVF}
 X\Big|_{\hyp}=N n\;,
\end{equation}
where $N$ is a function and $n=e_0
=({r^2\over\ell^2}+k)^{-1/2}\partial_t$ is the future-directed
$b$-unit normal to $\hyp$. We shall use the symbol $\cKS$ to
denote this set of Killing vectors.  The question then arises
whether one can extract out of \eq{toto}, with $X\in\cKS$, one or
more geometric invariants associated to $g$ along $\hyp$.  Another
way of stating this question is, essentially, whether the
integrals \eq{toto} are background independent. As discussed in
more detail below, every metric $g$ asymptotes many different
backgrounds of the form \eq{eq:a1} whenever it asymptotes one, and
it is not at all clear how these backgrounds relate to each other:
if the geometry of space-time does not sufficiently constrain the
set of allowed backgrounds \eq{eq:a1}, then the numbers obtained
from \eq{toto} could be completely unrelated to each other when
different backgrounds are chosen.  If this were the case, it would
appear questionable to associate physical meaning to the integrals
\eq{toto}. The purpose of this paper is to prove that, in several
cases of interest, geometric invariants can indeed be extracted
out of the integrals \eq{toto}.

The model problem of interest are space-times which are asymptotic
to anti-de Sitter space-time. In this context there exist several
alternative methods of defining mass --- using coordinate systems
\cite{BGH,GHHP83}, preferred foliations \cite{GibbonsGPI},
generalized Komar integrals \cite{Magnon}, conformal techniques
\cite{AshtekarDas,AshtekarMagnonAdS,Balasubramanian}, or
\emph{ad-hoc} methods \cite{AbbottDeser}; an extended discussion
can be found in \cite[Section 5]{ChruscielSimon}. We wish to
stress that the key advantage of the Hamiltonian approach is the
uniqueness of the candidate expression for the energy (which
follows from the fact that Hamiltonians are uniquely defined up to
a constant on each path connected component of the phase space),
and that no such uniqueness properties are known  in the
alternative approaches mentioned above (\emph{cf.}, however,
\cite{WaldZoupas,SilvaJulia:2000} for some partial results in the
``Noether charges'' framework). Now, independently of the question
of what is the ``correct'' candidate expression for the energy,
each of the expressions proposed in the existing literature
suffers from some ambiguities, so that the question of
well-posedness of the definition of mass as defined in those
papers arises as well. For instance, the Abbott-Deser
mass~\cite{AbbottDeser}, or the Hamiltonian mass of \cite{HT},
both suffer from precisely the same potential ambiguities as the
Hamiltonian mass analyzed in this paper. As shown in
Appendix~\ref{AAD}, under the asymptotic conditions considered in
our well-posedness results, the Hamiltonian mass defined by
\eq{toto} coincides with the Abbott-Deser one. Thus, one way of
interpreting our results is that we prove the existence of a
geometric invariant which can be calculated using Abbott-Deser
type integrals. As another example, we note the potential
ambiguity in the mass defined by the conformal methods
in~\cite{AshtekarDas,AshtekarMagnonAdS}, related to the
possibility of existence of conformal completions which are
\emph{not} smoothly conformally equivalent. The results proved
here can be used to show~\cite{ChHerzlich} that no such
completions exist, establishing the invariant character of the
definitions of~\cite{AshtekarDas,AshtekarMagnonAdS}.

We note that a similar problem for the ADM mass of asymptotically
flat initial data sets has been solved in
\cite{Bartnik:mass,ChErice} (see also \cite{Chmass}). Our
treatment here is a non-trivial adaptation to the current setup of
the methods of \cite{ChErice}. Some of the results proved here
have been independently observed in \cite{Wang}.

The detailed statements of our results in a general context are to
be found in the sections below, and will not be reproduced here.
We shall, instead, discuss the application of our results to two
families of examples:
\begin{enumerate}
\item Let $M$ be the $n-1$ dimensional sphere $^{n-1}S$ with the
  round metric $h$ of scalar curvature $(n-1)(n-2)$, while $b$ is the
  $n+1$ dimensional anti-de Sitter metric:
\begin{equation}
  \label{eq:a1a}
  b= -\left({r^2\over\ell^2}+1\right) dt^2 + {1\over{r^2\over\ell^2}+1 } dr^2
  +r^2 h\;,
\end{equation}
The space $\cKS$ of $b$-Killing vector fields normal to
$\hyp\cap\cU$ consists of vector fields $X(\lambda)$,
$\lambda=(\lambda^{(\mu)})\in \R^{n+1}$, which\footnote{We stress
that the index ${(\mu)}$ on
  $\lambda$ does not have anything to do with space-time;
  $\lambda^{(\mu)}$ is simply a coordinate on the $n+1$ dimensional
  vector space $\cKS$.  Similarly the Lorentz metric
  $\eta_{{(\mu)}{(\nu)}}$, which arises naturally on $\cKS$ from the
  construction here, does not have anything to do with the space-time
  metric $g$. To emphasize this we put brackets around the $\mu$'s.}
  on $\hyp$ take the form \eq{KVF} with $N=\lambda^{(\mu)}
  N_{(\mu)}$, where $$N_{(0)}=\sqrt{{r^2\over\ell^2}+1}\;, \qquad
  N_{(i)}={x^i\over\ell}\;,$$ and $x^i = r n^i$, $r$ being the
  coordinate which appears in \eq{eq:a1}, while
  $n^i\in{}^{n-1}S\subset\R^n$. The group $\mathit{Iso}(\hyp,b)$ of
  isometries $\Phi$ of $b$ which map $\hyp$ into $\hyp$ acts on $\cKS$
  by push-forward; in Appendix \ref{Aai1} we show for completeness the
  well known fact that for every such $\Phi$ there exists a Lorentz
  transformation $M:\R^{n+1}\to\R^{n+1}$ so that we have $$\Phi_*
  X(\lambda)= X( M\lambda)\;.$$ Letting ${\mathfrak g}_{(\mu)}$ be the
  $\mu$'th basis vector of $\cKS\approx\R^{n+1}$, $ {\mathfrak
  g}_{(\mu)}:=X(\lambda^{(\alpha)} =\delta^\alpha_\mu)$, we set
  $$m_{(\mu)}\equiv m\left(\hyp,g,b,{\mathfrak g}_{(\mu)}\right)\;;$$
  it follows that the number $$m^2(\hyp,g) = |\eta^{{(\mu)}{(\nu)}}
  m_{(\mu)} m_{(\nu)}|\;,$$ where
  $\eta^{{(\mu)}{(\nu)}}=\mathit{diag}(-1,+1,\cdots,+1)$ is the
  Minkowski metric on $\R^{n+1}$, is an invariant of the action of
  $\mathit{Iso}(\hyp,b)$.\footnote{This has been observed
  independently by X.~Wang~\cite{Wang} in, however, a considerably
  less general setting.} Further, if we define $m(\hyp,g)$ to
  be positive if $m^\mu$ is spacelike, while we take the sign of
  $m(\hyp,g)$ to coincide with that of $m_0=-m^0$
  if $m^\mu$ is timelike or
  null, then $m(\hyp,g)$ so defined is invariant under the action of
  the connected component $\mathit{Iso}_0(\hyp,b)$ of the identity in
  $\mathit{Iso}(\hyp,b)$. We show in detail in Section~\ref{Sgc} that
  $m(\hyp,g)$ is independent of the background metric chosen to
  calculate the integrals \eq{toto}, provided that the fall-off
  conditions of Theorem~\ref{T0} and Theorem~\ref{TP1} hold, which
  justifies the notation. The number $m(\{t=0\},g)$ so defined
  coincides with the mass parameter $m$ of the Kottler
  metrics\footnote{The Kottler metrics, published in 1918
  \cite{Kottler}, are also known as the ``Schwarzschild -- de Sitter
  metrics''.}  \begin{eqnarray}\label{Kottler} g & = & -\left(1 -
  \frac {2m}r + {r^2\over \ell^2}\right) dt^2 + \left(1 - \frac {2m}r
  + {r^2\over \ell^2}\right)^{-1} dr^2 + r^2 h \;,
\end{eqnarray}
where $h=d\theta^2+\sin^2\theta d\varphi^2$. Similarly
$m(\{t=0\},g)$ is proportional to the parameter $\eta$ which
occurs in the $(n+1)$-dimensional generalizations of the Kottler
metrics ({\em cf., e.g.,} \cite{HorowitzMyers})
 \begin{eqnarray}\label{Kottler2}
  g & = & -\left(1 - \frac {2\eta}{r^{n-2}} +  {r^2\over \ell^2}\right) dt^2 +
  \left(1 - \frac
  {2\eta}{r^{n-2}} +  {r^2\over \ell^2}\right)^{-1} dr^2 + r^2 h \;,
\end{eqnarray}
with $h$ --- a round metric on a $n-1$ dimensional sphere of
scalar curvature $(n-1)(n-2)$.

Some further global geometric invariants of the metrics asymptotic
to the backgrounds \eq{Kottler2} are discussed in
Section~\ref{Sgc}.
\item  Let $M$ be a compact $n-1$ dimensional manifold
  with a metric $h$ of constant scalar curvature and with non-positive
  Ricci tensor, and let $b$ take the form
\begin{equation}
  \label{eq:a1c}
  b= -{1\over a^2(r)} dt^2 + a^2(r) dr^2 +r^2 h\;,
\end{equation}
$h$ being $r$-independent, as before. We show (see
Proposition~\ref{PA1}, Appendix~\ref{Aai2}) that for such metrics
the space of $b$-Killing vector fields normal to $\hyp$ consists
of vector fields of the form
\begin{equation}
  \label{eq:kvh}
X(\lambda)= \lambda\partial_t\;, \quad\lambda\in\R\;.
\end{equation}
The discussion in Section~\ref{Sgc} shows that $$m(\hyp,g)\equiv
m(\hyp,g,b,X(1))$$ is background independent, hence a geometric
invariant.  Some other geometric invariants can be obtained from
the integrals \eq{toto} when Killing vectors which are not
necessarily normal to $\hyp$ exist, using invariants of the action
of the isometry group of $b$ on the space of Killing vectors.  If
the Ricci tensor of $\mn$ is strictly negative no other Killing
vectors exist, \emph{cf.\/} Appendix~\ref{Aai2}. On the other
hand, if $h$ is a flat torus, then each $h$--Killing vector
provides a geometric invariant via the integrals \eq{toto},
provided that those converge and that the fall-off conditions of
Theorem~\ref{TP1} are met (this will be the case if,
\emph{e.g.,\/} Equations~\eq{C4}-\eq{C5} hold).
\end{enumerate}

The number $m(\hyp,g)$ defined in each case above is our proposal
for the geometric definition of total mass of $\hyp$ in $(\cM,g)$.

The results described above can be reformulated in a purely
Riemannian context, this will be discussed elsewhere
\cite{ChHerzlich}. The extension of the results proved here to
hyperboloidal hypersurfaces in Minkowski space-time, that  leads
to a geometric definition of the Trautman-Bondi mass, requires a
considerable amount of work and will be discussed
elsewhere~\cite{ChJTB}. Let us simply mention that if  the metric
of a hyperboloidal hypersurface in asymptotically Minkowskian
space-times satisfies the fall-off conditions here then its
Trautman-Bondi mass coincides with the Hamiltonian one. More
general statements require care.

It is natural to study the behaviour of the mass when $\hyp$ is
allowed to move in $\cM$. A partial answer to this question is
given in Theorem~\ref{TP1} below. A complete answer would require
establishing an equivalent of Theorem~\ref{T1} in a space-time
setting. The difficulties that arise in the corresponding problem
for asymptotically Minkowskian metrics \cite{Chmass} suggest that
this might be a considerably more delicate problem, which we plan
to analyze in the future. It should be stressed that this problem
mixes two different issues, one being the potential background
dependence of \eq{toto}, another one being the possibility of
energy flowing in or out through the timelike conformal boundary
of space-time.

This paper is organized as follows: In Section~\ref{S2} we present
conditions which guarantee convergence of the mass integrals
\eq{toto}, see Theorem~\ref{T0}. We also show that the integrals
\eq{toto} are invariant (Theorem~\ref{TP1}) or covariant
(Lemma~\ref{L1}) under a class of well-controlled coordinate
transformations, consisting of symmetries of the background, and
of certain generalizations of the usual ``supertranslations'' that
occur in the asymptotically flat case.  Section~\ref{Sai} contains
the proof of the asymptotic symmetries theorem, Theorem~\ref{T1},
which is the key result in this work.  In this theorem we show
that the coordinate transformations allowed by our conditions are
compositions of those considered in Section~\ref{S2}. In
Section~\ref{Sgc} we apply the previous results to the
construction of global geometric invariants in a reasonably
general setting.  In Appendix~\ref{AHam} the Hamiltonian approach
to the definition of mass  is examined in our context.
Appendix~\ref{Aiso} contains some results on Killing vectors which
are needed in the body of the paper.

\section{Convergence, covariance under well behaved coordinate
  transformations}\label{S2}
\newcommand{\zero}{0} \newcommand{\one}{1} \newcommand{\backg}{b}
\newcommand{\A}{A} \newcommand{\ta}{a} \newcommand{\ha}{a}
\newcommand{\hc}{c} \newcommand{\gf}{f} \newcommand{\ce}{\beta} Let us
start by establishing convergence of the mass integrals \eq{toto}
--- this involves setting up appropriate boundary conditions on
$g$.  Let, thus, $g$ and $b$ be two metrics on a set $\{R_0\le r
<\infty\;, \; (v^A)\in \mn\}$, let $e_a$ be an orthonormal frame
for $b$, set
\begin{equation}
  \label{mas3}
  e^{\mu\nu}\equiv g^{\mu\nu}-b^{\mu\nu}
\; ,
\end{equation} and let  $e^{\ha b}\equiv g(\theta^a,\theta^b)-\eta^{ab}$
denote the coefficients of $e^{\mu\nu}$ with respect to the frame
$\theta^a$ dual to the $e_a$'s:
$$e^{\mu\nu}\partial_\mu\otimes \partial_\nu=e^{\ha\hc}e_{\ha}\otimes
e_{\hc}\;.$$ Here $\eta^{ab}=\mathrm{diag}(-1,+1,\cdots,+1)$. We
stress that we do {\em not} assume existence of global frames on
the asymptotic region: when $M$ is not parallelizable, then any
conditions on the $e^{ab}$'s, {\em etc.} assumed below should be
understood as the requirement of {\em existence of a covering of
$M$ by a finite number of  open sets ${\mycal O}_i$ together with
frames defined on $[R_0,\infty)\times {\mycal O}_i$ satisfying the
conditions spelled out above.}  The ``matter energy-momentum
tensor'' $T^\lambda{}_\kappa$ is defined as \be\label{N.5g2} 8\pi
T^\lambda{}_\kappa:= R^\lambda{}_\kappa -\frac 12
g^{\alpha\beta}R_{\alpha\beta} \delta ^\lambda_\kappa +\Lambda
\delta ^\lambda_\kappa\;. \ee In our first result we assume for
simplicity\footnote{Using \Eq{C3}, Appendix~\ref{AHam}, it is
straightforward to obtain results similar to Theorem~\ref{T0}
without the hypothesis that $b$ is Einstein. Similarly, the
hypothesis that $X$ is a Killing vector field can be relaxed using
the calculations of \cite[Appendix~B]{ChAIHP}; {\em cf.\/} also
\cite[Section~5.1]{CJK}.} that $b$ is Einstein, that is, $b$
satisfies \Eq{N.5g2} with $T^\lambda{}_\kappa=0$, with a
cosmological constant $\Lambda$ the sign of which is irrelevant
for the theorem that follows:
\begin{Theorem}\label{T0}
  Let $X$ be a Killing vector of an Einstein metric $b$,
set
  \be\label{c0}|X|^2\equiv \sum_a |X^a|^2\;, \quad |\znabla X|^2\equiv
\sum_{a,b} |\znabla _bX^a|^2\;,\quad |J|^2=
\sum_{b}|T^0{}_{b}|^2\;
\ee 
where $\znabla$ is the covariant derivative of $b$; the indices
here refer to a $b$-orthonormal frame such that $e_0$ is normal to
the hypersurface $t=0$.
Suppose that $\lim_{r\to\infty} e^{ab}=0$ and that
\begin{eqnarray}\nonumber
  \int_{\{r\ge R_0\}} \left\{|X|\left( |J|+{|\Lambda|} |e^{ab}b_{ab}| +
      \sum_{a,b,c} e_a(e^{bc})^2     +\sum_{b,c}
    |e^{bc}|^2   \right) \right.
\\ \left. + |\znabla X| \sum_{a,b,c,d,e}
    |e_a(e^{bc})| \; |e^{de}| \right\} \; d\mu_b < \infty\;,
  \label{c1}\end{eqnarray}
where $d\mu_b\equiv \sqrt{\det b_{ij}}dr\; dv^2\ldots dv^{n}$ is
the Riemannian measure induced on $\{t=0\}$ by $b$. Then the
right-hand-side of \Eq{toto}, understood as the limit  as $R\to
\infty$ of integrals over the sets $\{r=R, t=0\}$, exists and is
finite.
\end{Theorem}

\remark We note that the somewhat unexpected restriction on
integrability (for $\Lambda \ne 0$) of $e^{ab}b_{ab}$ arises also
in the requirement of a well defined generalized Komar mass for
static asymptotically anti-de Sitter metrics
\cite{Magnon,ChruscielSimon}.

\medskip

\proof We have
\begin{eqnarray}
   \int_{\{r=R\}}
 \ourU^{\alpha\beta}dS_{\alpha\beta}=  2\int_{\{R_0\le r\le R\}}
 \znabla_\beta \ourU^{\alpha\beta}dS_{\alpha}  + \int_{\{r=R_0\}}
 \ourU^{\alpha\beta}dS_{\alpha\beta}\;.
\label{C2}
\end{eqnarray}
A formula for the volume integrand in \Eq{C2} is given in \Eq{C3},
Appendix~\ref{AHam}. Clearly conditions \eq{c1} guarantee
convergence of that volume integral to a finite value when $R$
tends to infinity. \qed

In the remainder of the paper we will only consider background
metrics of the form
\begin{equation}
  \label{A1}
  b= -{ a^{-2}(r)} dt^2 + a^2(r) dr^2 +r^2 h\;, \qquad
  h=h_{AB}(v^C)dv^A dv^B\;,
\end{equation}
where $h$ is a Riemannian metric on a $(n-1)$-dimensional compact
manifold $\mn$. The condition that $b$ is Einstein will not be
made unless explicitly stated otherwise. Let
$$\theta^0 = {1\over a} dt\;, \quad \theta^1= a dr\;, \quad \theta^A = r
\alpha^A\;,$$ where $\alpha^A$ is an $h$-orthonormal coframe. We
let $e_a$ be the frame dual to $\theta^a$,
\begin{eqnarray}
  \label{frame}
e_0 = a \partial_t\;, \qquad e_1 = {1 \over a } \partial _r
\quad
  e_{\A} = \frac 1r \ce_{\A}\;,
\end{eqnarray}
so that $\ce_A$ is a $h$-orthonormal frame on $(M,h)$. As an
application of Theorem~\ref{T0}, consider, first, the Killing
vector field $X=\partial_t=(1/a)e_0$ and suppose that
\begin{equation}
  \label{A1.0} a(r)={\ell\over r } +
o(r)\;,\quad e_1(a)= -{1\over r } + o(r)\;, \end{equation} for
some constant $\ell$. We then have $|X|\approx \ell|\znabla
X|/\sqrt 2\approx{r/ \ell}$ and $d\mu_b= r^{n-2}\sqrt{\det
h}\,dr\, dv^2\ldots dv^{n}$.
  When $g$ and $b$ are Einstein, the condition \eq{c1} will be
  satisfied if
\begin{equation}
  \label{C4}
  e^{ab}= O(r^{-\beta})\;, \quad e_a(e^{bc}) = O(r^{-\beta})\;, \quad
  b_{ab}e^{ab}= O(r^{-\gamma})\;,
\end{equation}
with
\begin{equation}
  \label{C5}
  \beta>n/2\;, \qquad \gamma > n\;.
\end{equation}
(We note that the generalized $n+1$ dimensional Kottler metrics
\eq{Kottler2} satisfy \eq{C4} with $\beta=n$, and with
$\gamma=2n$.) An identical convergence analysis applies for all
``rotational'' $b$-Killing vector fields $X^A(v^B)\partial_{A}$
(whenever occurring) for all the metrics \eq{A1}-\eq{A1.0}, as
well as all the remaining Killing vectors for the
$(n+1)$-dimensional anti-de Sitter metric (listed in
Appendix~\ref{Aai1}), showing that all the corresponding charges
are finite when the conditions \eq{C4}-\eq{C5} hold. Surprisingly,
in retrospect the analysis in the case $\Lambda\ne 0$ turns out to
be simpler than that for the asymptotically Minkowskian case,
where $\Lambda = 0$: in the latter case the requirement of
convergence of angular momentum or of boost integrals imposes more
stringent conditions on the metric than that of convergence of the
energy-momentum integrals.

The conditions presented above are sufficient, but certainly not
necessary, for convergence of the integrals \eq{toto}: indeed, the
metric considered in Proposition~\ref{P2} below has a convergent
mass integral, but the conditions of Theorem~\ref{T0} fail to
hold. However, there is a potential essential ambiguity in the
definition of the integrals \eq{toto}, which  we will describe
now. Proposition~\ref{P2} below then shows that \eq{C4}-\eq{C5}
are essentially sharp, if one requires that the integrals
\eq{toto} are convergent and {\em background-independent}.

The ambiguity in the integrals \eq{toto} arises as follows: to
define those integrals we have fixed a model background metric $b$
with the corresponding coordinate system $(t,r,v^A)$ as in
\eq{A1}, as well as an orthonormal frame as in \eq{frame}. Once
this has been done, let $g$ be any metric such that the frame
components $g^{ab}$ of $g$ tend to $\eta^{ab}$ as $r$ tends to
infinity in such a way that the integrals $m(\hyp,g,b, X)$ given
by \eq{toto} (labeled by the background Killing vector fields $X$
or perhaps by a subset thereof) converge.  Consider another
coordinate system $(\hat t,\hat r,\hat v^A)$ with the associated
background metric $\hat b$:
\begin{equation}
  \label{A1hat}
  \hat b= -{1\over a^2(\hat r)} d\hat t^2 + a^2(\hat r) d\hat r^2 +\hat r^2 \hat h\;, \qquad
  \hat h=h_{AB}(\hat v^C)d\hat v^A d\hat v  ^B\;,
\end{equation}
together with an associated frame $\hat e^a$,
\begin{eqnarray}
  \label{hframe}
\hat e_0 = a(\hat r) \partial_{\hat t}\;, \qquad \hat e_1 = {1
\over
  a(\hat r)  } \partial _{\hat r}
\quad
  \hat e_{\A} = \frac {1 }{\hat r} \hat{\ce}_{\A}\;,
\end{eqnarray}
and suppose that in the new hatted coordinates the integrals
defining the charges $m(\hhyp,g,\hat b, \hat X)$ converge again.
An obvious way of obtaining such coordinate systems is to make a
coordinate transformation
\begin{equation}
  \label{eq:A5}
  t\to \hat t=t+\delta t\;,\qquad r\to \hat r=r+\delta r\;,\qquad
v^A\to \hat v^A=v^A+\delta v^A\;,
\end{equation}
with $(\delta t, \delta r,\delta v^A)$ decaying sufficiently fast,
as \emph{e.g.\/} in the statement of Theorem~\ref{TP1} below.
(However, we do not know \emph{a priori\/} that the hatted
coordinates are related to the unhatted one by the simple
coordinate transformation \eq{eq:A5} with $(\delta t, \delta
r,\delta v^A)$ decaying as $r\to\infty$, or behaving in some
controlled way --- the behaviour of $(\delta t, \delta r,\delta
v^A)$ could be very wild.) The question then arises, how do the
$m(\hhyp,g,\hat b, \hat X)$'s relate to the $m(\hyp,g,b, X)$'s. A
geometric definition of mass should be coordinate-independent,
therefore one would like to have a simple relation between those
integrals.

At this point it is worth recalling that there exist several
expressions for mass alternative to \eq{toto}, which might or
might not coincide with each other when the decay of the metric is
too slow. For example, we show in Appendix~\ref{AAD} that
\eq{toto} coincides with the Abbott-Deser \cite{AbbottDeser} mass
for all metrics satisfying the decay conditions \eq{C4}-\eq{C5}
for Killing vectors such that $|X|=O(r)$ with, say, $a(r)$ as in
\Eq{eq:a1}. Now, if $X=\partial_t$, for background metrics of the
form \eq{eq:a1}, in space-times of dimension $4$, the integral
defining the Abbott-Deser $m_{AD}$ can be written in a
particularly simple form \cite{ChruscielSimon}
 \begin{eqnarray}
   m_{AD}(\{t=0\},g, b, \partial_t)
=\lim_{R\to\infty} \frac {R^3}
   {16\pi\ell^2}\int_{\Sigma\cap\{r=R\}} \left(r\sum_A{\frac {
          \partial e^{AA}}{\partial r}}-2e^{\one\one}
   \right) d^2\mu_{h}\;.
   \label{massequation}
 \end{eqnarray}
 Generalizing an argument of \cite{DenisovSolovev}, we show that if
 the decay conditions in \eq{C4} are too weak then one can obtain
 essentially any value of $m_{AD}(\hhyp,g,\hat b, \hat X)$ by performing
 coordinate transformations of the form \eq{eq:A5}.  We do this
 explicitly in $n=3$, the same argument applies in any dimension $n$:

\begin{Proposition}\label{P2}
  Let the physical metric $g$  equal the background metric $b$, and
  let $\{r,v^A\}$ be coordinates so that $b$ takes the form
  \eq{eq:a1a}.  Consider a new set of coordinates defined as
\be\label{ct1} \hat{r} = r+ \frac{\ac}{r^{1/2}} \;,\quad
\hat{v}^A= v^{A}\;,\ \ee
 where $\ac$ is a constant. (This leads to $\hat e^{ab}=O(r^{-3/2})$.)
If $\ac\ne 0$ then the mass $m_{AD}(\{t=0\},g,\hat b,
\partial_t) $ of $g$ with respect to the background metric $\hat b$
defined by the coordinates $\{\hat{r},\hat{v}^A\}$ does not
vanish.
\end{Proposition}

\proof First notice that this transformation satisfies
$\frac{\delta r}{r} = O\left( r^{-3/2}\right)$.  Then, by
straightforward computations one has, assuming without loss of
generality that  $\ell=1$,
$$
\hat{e}_{1} = \left[ 1+ \frac{3\ac}{2r^{3/2}} +
\frac{3\ac^2}{4r^3}  + O( {r^{-7/2}})\right] e_1\;,
$$
$$
\hat{e}_{A} = \left[ 1- \frac{\ac}{r^{3/2}} + \frac{\ac^2}{r^3}
 +  O( {r^{-7/2}})\right] e_A\;,
$$
and so,
$$
\hat{e}^{11} =   \frac{3\ac}{r^{3/2}} + \frac{15 \ac^2}{4r^3} +O(
{r^{-7/2}})\;,
$$
$$
\sum_A{e}^{AA} =  -\frac{4\ac}{r^{3/2}} + \frac{6\ac^2}{r^3} +  O(
{r^{-9/2}})\;.
$$
Hence
$$
r\sum_A\partial_r({e}^{AA}) - 2 {e}^{11} = -\frac{51 \ac^2}{2
  r^3} +  O( {r^{-7/2}})\;,
$$
and the result follows. \qed

While the above shows that the Abbott-Deser mass ceases to be well
defined below the threshold $o(r^{-3/2})$ in dimension $3+1$, this
still leaves open the unlikely possibility that the Hamiltonian
mass \eq{toto} could be well defined. In order to see that this is
not the case let us, first, calculate the mass integrand for
metrics of the form \bel{acom} g=  -\mu
\left(\frac{r^2}{\ell^2}+k\right) dt^2 +\displaystyle
\frac{\nu}{\frac{r^2}{\ell^2}+k} dr^2 + \sigma r^2
h_{AB}dv^Adv^B\;, \ee where $\mu$, $\nu$ and $\sigma$ are
arbitrary functions. One finds
\begin{eqnarray}
\ourU^{tr}
&=&\frac{r^{n}\sigma^{(n-1)/2}\sqrt{\det{h_{AB}}}}{16\pi
\ell^2\sqrt{\mu\nu}} \left\{
\frac{2\sqrt{\mu\nu}}{\sigma^{(n-1)/2}}-
(\mu+\nu)\right.\nn\\
&&+ \left. (n-1)\left(1+\frac{\ell^2k}{r^2}\right)\mu
\left[\frac{1}{\sigma}\left(r\frac{\partial \sigma}{\partial r}
-\nu\right)+ 1\right]\right\}\;.\label{acomm}
\end{eqnarray}
Suppose that ---in space-time dimension $n+1$ ---  $g$ is the
metric $b$ expressed in a hatted coordinate system $(\hr,\hv^A)$,
and consider the coordinate transformation
$$ \hat{r} = r+ \frac{\ac}{r^{n/2-1}} \;,\quad \hat{v}^A= v^{A}\;,
$$
where $\ac$ is a constant. The metric $g$, when expressed in the
unhatted coordinates $(r,v^A)$, satisfies \eq{C4} with
$\beta=\gamma=n/2$, and is of the form \eq{acom} so that
\eq{acomm} applies. A {\sc Mathematica} calculation then shows
that $g$ has a mass integral \eq{toto} with respect to the
unhatted background $b$ equal to
$$
\frac{\text{Vol}_h(M)}{8\pi\ell^2} \left(n+{n^2\over
    8}\right)\ac^2\;,$$ which is non-zero for any $\ac\ne 0$ and for
    any $n\in\N$. Here $\text{Vol}_h(M)$ is the $n-1$-dimensional volume of $M$
    ---
    area if $n-1=2$ --- with respect to the metric $h$.

  The coordinate transformation \eq{ct1} is not yet as good
as one would wish, because it leads --- in space dimension three
-- to coordinates in which the trace of the metric $e^{ab}b_{ab}$
is $O(r^{-n/2})$, quite a bit above the threshold $r^{-n}$ set
forth in \Eqs{C4}{C5}. We note that the change of coordinates
\eq{ct1} accompanied by a further time redefinition (which clearly
does not change the mass as given by \Eq{massequation})
$$ \bar t =
t(1+cr^{-3/2})\;,$$ with an appropriate choice of the constant
$c$, will lead to a metric which at $\bar t = t=0$ satisfies
\begin{equation}
  \label{C4.}
  e^{ij}= O(r^{-3/2})\;, \quad e_k(e^{ij}) = O(r^{-3/2})\;, \quad
  b_{ab}e^{ab}= O(r^{-3})\;,
\end{equation}
where the indices $i,j$ run from $0$ to $3$. Note that the above
fall-off conditions will not hold for some of the $e^{0a}$'s, and
for some $e_0$ derivatives of the $e_{ab}$'s, but this turns out
irrelevant for the problem at hand: the new hypersurface $\bar
t=0$ coincides with the previous one, therefore its extrinsic
curvature will not change. One can check~\cite{ChHerzlich} that
 --- similarly to the ADM case
--- conditions on the induced metric on the surface $t=0$ and on
its extrinsic curvature are sufficient for a well defined notion
of mass, so that the result in~\cite{ChHerzlich} complete the
proof of sharpness of the condition on $\gamma$ in \eq{C5}.

Let us show that the decay rates \eq{C4}-\eq{C5} guarantee
non-occurrence of the behaviour exhibited in Proposition~\ref{P2}:

\begin{Theorem}\label{TP1}
 Consider an $n+1$ dimensional space-time
$(\cM, g)$, and let $b$ and $\hat{b}$ be two background metrics of
the form \eq{A1} and \eq{A1hat}, with $a(r)$ as in \Eq{eq:a1}, in
coordinates $\{t,r,v^{A}\}$ and $\{\hat{t},\hat{r},\hat{v}^{A}\}$
with ranges covering $\{r\ge R_0\}\times\mn$ and $\{\hat r\ge \hat
R_0\}\times\mn$ for some $R_0,\hat R_0\in \R$. Suppose that $b$
satisfies the vacuum Einstein equations with a negative
cosmological constant, that the conditions of Theorem \ref{T0}
hold both for the hatted and unhatted coordinates, and that we
have
\begin{eqnarray}
 e^{ab}= o(r^{-n/2})\;, \quad e_c(e^{ab})= o(r^{-n/2})\;.
   \label{eq:deccond}
\end{eqnarray}
 Let $X
=X^a({t},{r},{v}^A)e_a\in \cK$ be a Killing vector field of the
metric $b$ satisfying
\begin{eqnarray}
|X| + |\znabla X| = O(r)\;,   \label{eq:deccond1}
\end{eqnarray}
and let $\hat{X}=X^a(\hat{t},\hat{r},\hat{v}^A)\he_a \in \hcK$ be
its hatted counterpart (with the $\he_a$ components of $\hat X$
given by the {\em same} functions $X^a$ of the hatted variables as
the $e_a$ components of $X$ in the unhatted variables). Let $\hyp$
and $\hhyp$ be the hypersurfaces given by $t=0$ and $\hat{t} =0$
respectively. If the coordinate transformation satisfies
$$
\hat{t} = t + o(r^{-(1+n/2)})\;,\qquad e_a(\hat{t}) =
\ell\,\delta_a^0 + o(r^{-(1+n/2)})\;,
$$
$$
\hat{r} = r + o(r^{1-n/2})\;,\qquad e_a(\hat{r}) =
\frac{\delta_a^1}{\ell} + o(r^{1-n/2})\;,
$$
\be \hat{v}^A = v^A + o(r^{-(1+n/2)})\;,\qquad e_a(\hat{v}^A) =
\delta_a^A + o(r^{-(1+n/2)}) \;, \label{coordtrP}\ee  then
$$m(\hyp,g,b,X)=m(\hhyp,g,\hat{b},\hat{X})\;.$$
\end{Theorem}

\proof  The idea of the proof is to compute the background metric
$\hat{b}$ in a frame related to the unhatted coordinates,
obtaining an expression in terms of $b$ plus correction terms.
Then, we compute (\ref{Fsup2new}) for $\hat{b}$, similarly
obtaining an expression in terms of (\ref{Fsup2new}) for $b$ plus
additional terms. We show that these terms integrate to zero, up
to terms vanishing in the limit as $r$ tends to infinity, keeping
thus the mass invariant.

In terms of $\delta t$, $\delta r$, and $\delta v^A$ defined as in
\Eq{eq:A5}, the decay conditions \eq{coordtrP} imply
\begin{equation}
\label{fc-dx} \sqrt{k+r^2/\ell^2} \, \delta t =o(r^{-n/2})\;,
\quad \frac{\delta r}{ \sqrt{k+r^2/\ell^2}} = o(r^{-n/2})\;, \quad
r \delta v^A = o(r^{-n/2})\;.
\end{equation}
From \Eq{hframe} one finds the following relation between the
hatted and non-hatted coframes
 $$
\hat{\theta}^0 =  \left(1+\frac{r\delta r}{{r^2+ k
\ell^2}}\right)\theta^0 +
 {\sqrt{k+r^2/\ell^2}}d(\delta t)  +
o(r^{-n})\;,
$$
 $$
\hat{\theta}^1 = \left(1-\frac{r\delta r}{{r^2+ k \ell^2}}\right)
\theta^1 +
 \frac 1{\sqrt{k+r^2/\ell^2}}d(\delta r)  +
o(r^{-n})\;,
$$
\be\label{tettr} \hat{\theta}^A = \theta^A + \frac{\delta r}{r}
\theta^A + r \left\{ \frac{\partial \alpha^A{}_B}{\partial
v^C}\,\delta v^C \, dv^B + \alpha^A{}_B \,d(\delta v^B) \right\} +
o(r^{-n})\;. \ee
 where $\alpha^A$ ia a co-frame on $M$ dual to $\beta_A$, and
we have introduced the notation $\alpha^A = \alpha^A{}_B dv^B$,
with the index $A$ being a tetrad index, while the index $B$ is a
coordinate index. All the terms denoted by $o(r^{-n})$ above have
$o(r^{-n})$ coefficients when expressed in terms of the $\theta^a$
frame. Actually, the term in the curly brackets in the right hand
side of the last equation gives a clue to a convenient way of
writing these equations, since it can be rewritten as
$\lie_{\delta v^B
\partial/\partial v^B }\alpha^A$, where $\lie$ denotes a Lie
derivative; in order to justify such a procedure, we use the
following artifact: As explained in \cite[Section~4]{CT93},
embedding $\mn$ in $\R^{2(n-1)}$ and extending the metric
appropriately, we can without loss of generality assume that the
coordinates $v^A$ and the frames $\theta^A$ are globally defined
on $\mn$. We then set $$ \zeta = \delta t \frac{\partial}{\partial
t} + \delta r \frac{\partial}{\partial r} + \delta v^A
\frac{\partial }{\partial v^A}\;; $$ one sees from (\ref{fc-dx})
that the components of $\zeta$ in the $e_a$ frame are all of the
same order $o(r^{-n/2})$, and one can check that \Eq{tettr} for
the hatted tetrad reduces to $$ \hat{\theta}^a = \theta^a +
\lie_{\zeta}\theta^a + o(r^{-n})\;;
$$ we note that $\lie_{\zeta}\theta^a = o(r^{-n/2})$. Let us write
$$\hat{e}_a = e_a + \delta e_a\;;$$ one verifies that the tetrad
components of $\delta e_a$ are $o(r^{-n/2})$, and from the
condition $\theta^a(e_b) = \delta^a_b$ and its hatted analogue one
has
\begin{eqnarray*} \theta^a(\delta e_b) &=&- \lie_{\zeta}\theta^a(e_b)
+ o(r^{-n})
\\ & = &
- \underbrace{\lie_{\zeta}\left(\theta^a(e_b)\right)}_0+
\theta^a(\lie_{\zeta}e_b) + o(r^{-n})
\\ & = &
 \theta^a(\lie_{\zeta}e_b) + o(r^{-n})
\;,
\end{eqnarray*}
which shows that
$$
\hat{e}_a = e_a + \lie_{\zeta}e_a + o(r^{-n})\;,
$$
with $\lie_{\zeta}e_a = o(r^{-n/2})$. This looks to leading order
like a change of tetrad under an infinitesimal transformation, but
we emphasize that \emph{we are not} assuming that the
transformation is infinitesimal. Denoting $\{x^{\mu}\} =
\{t,r,v^A\}$, and using $\hat{b}^{\mu\nu} = \eta^{ab}
\hat{e}_a{}^{\mu}\hat{e}_b{}^{\nu}$, and $b^{\mu\nu} = \eta^{ab}
e_a{}^{\mu}e_b{}^{\nu}$, we obtain
\begin{equation}
\label{b-freedom} \hat{b}^{\mu\nu} = b^{\mu\nu} +
\lie_{\zeta}b^{\mu\nu} + \delta b^{\mu\nu}\;,
\end{equation}
with $$\delta b^{ab}\equiv \delta
b^{\mu\nu}\theta^a{}_{\mu}\theta^b{}_{\nu} = o(r^{-n})\;.$$ The
expression \eq{b-freedom} is the first step to compute the change
in the integrand of (\ref{toto}). The next step is to rewrite
(\ref{Fsup2new}) in the following more convenient way $$
\ourU^{\alpha\beta} = \frac{3}{8\pi} \sqrt{|\frg|}
g_{\gamma\sigma} g^{\kappa[\alpha} X^{\beta}
\zn_{\kappa}g^{\gamma]\sigma} + \frac{1}{8\pi} \left(\sqrt{|\frg|}
g^{\kappa[\alpha}b^{\beta]\gamma} - \sqrt{|\fb|}
b^{\kappa[\alpha}b^{\beta]\gamma}\right)
b_{\gamma\sigma}\zn_{\kappa}X^\sigma\;, $$ with
$\frg=\det(g_{\mu\nu})$, and $\fb=\det(b_{\mu\nu})$. Let
$\widehat{\ourU}^{\alpha\beta}$ be defined as the expression above
with $b$ and $X$ replaced with $\hat{b}$ and $\hat X$; from
\Eq{b-freedom} we obtain
\begin{eqnarray*}
\widehat{\ourU}^{\alpha\beta} -\ourU^{\alpha\beta} &=&
-\frac{\sqrt{|\fb|}}{8\pi}  \left[
3\delta_{\lambda\nu\mu}^{\alpha\beta\gamma} X^{\mu}
b^{\kappa\lambda} b_{\gamma\sigma} (\zn_{\kappa}\zn^{\nu}
\zeta^{\sigma} + \zn_{\kappa}\zn^{\sigma}\zeta^{\nu}) \right] \\
&& - \frac{\sqrt{|\fb|}}{8\pi} \left[
b^{\kappa[\alpha}b^{\beta]\gamma} \zn_{\mu}\zeta^{\mu} -
b^{\gamma[\beta}\zn^{|\kappa|}\zeta^{\alpha]} - b^{\gamma[\beta}
\zn^{\alpha]}\zeta^{\kappa} \right] \zn_{\kappa}X_{\gamma} +
\delta\ourU^{\alpha\beta}\;,
\end{eqnarray*}
with $$\delta \ourU^{ab}\equiv\delta \ourU^{\alpha\beta}
\theta^a{}_{\alpha} \theta^b{}_{\beta} = o(r^{-n})\;.$$ We have
used the fact that $\lie_{\zeta} b^{\alpha\beta}= -2
\zn^{(\alpha}\zeta^{\beta)}$, and  that $\sqrt\frg =\sqrt\fb(1 +
\zn_{\mu}\zeta^{\mu} - e_\mu{}^\mu/2) + o(r^{-n})$.

The idea now is to write the right hand side above as a total
divergence of a totally antisymmetric tensor density. The first
term at the right hand side above can be written as
\begin{eqnarray*}
\lefteqn{ 3\delta_{\lambda\nu\mu}^{\alpha\beta\gamma} X ^{\mu}
b^{\kappa\lambda} b_{\gamma\sigma} (\zn_{\kappa}\zn^{\nu}
\zeta^{\sigma} + \zn_{\kappa}\zn^{\sigma}\zeta^{\nu})  =
 3 \zn_{\gamma} \left[
\delta^{\alpha\beta\gamma}_{\lambda\nu\mu} X
^{\mu}\zn^{\lambda}\zeta^{\nu}\right] } && \\
&&+\zRm_{\gamma\rho}{}^{\alpha\beta} X ^{\gamma}\zeta^{\rho}  + 2
X ^{[\beta}\zRm^{\alpha]}{}_{\rho} \zeta^{\rho} - (\zn_{\gamma}X
^{\beta}) \zn^{[\gamma}\zeta^{\alpha]} +(\zn_{\gamma}X ^{\alpha})
\zn^{[\gamma}\zeta^{\beta]} \;.
\end{eqnarray*}
Since $\zn_{\alpha}X _{\beta}$ is antisymmetric, one obtains
\begin{eqnarray*}\hat{\ourU}^{\alpha\beta} -\ourU^{\alpha\beta} &=&
-\frac{\sqrt{|b|}}{8\pi} \left\{3\zn_{\gamma} \left[ \,
  \delta^{\alpha\beta\gamma}_{\lambda\nu\mu} X
  ^{\mu}\zn^{\lambda}\zeta^{\nu}\right]
+\zRm_{\gamma\rho}{}^{\alpha\beta} X ^{\gamma}\zeta^{\rho} \right.\\
&& \left.+ 2 X ^{[\beta}\zRm^{\alpha]}{}_{\rho} \zeta^{\rho} + 3
(\zn^{[\alpha}X ^{\beta})(\zn_{\gamma}\zeta^{\gamma]})\right\}+
\delta \ourU^{\alpha\beta}\;.
\end{eqnarray*}
Finally, with the remaining terms we construct the divergence of a
totally antisymmetric tensor, as follows: $$ 3
(\zn_{\gamma}\zeta^{[\gamma})(\zn^{\alpha}X ^{\beta]}) = 3
\zn_{\gamma}\left[\zeta^{[\gamma}(\zn^{\alpha}X ^{\beta]})\right]
- \zeta^{\gamma}\zn_{\gamma}\zn^{\alpha}X ^{\beta} - 2
\zeta^{[\alpha}\zRm^{\beta]}{}_{\rho} X ^{\rho}.
$$ Recalling that a Killing vector satisfies $\zn_{\alpha}\zn_{\beta}
X _{\gamma} = \zRm^{\rho}{}_{\alpha\beta\gamma} X _{\rho}$, and
that $\zRm_{\alpha\beta} = 2\Lambda b_{\alpha\beta}/(n-1)$, one
finds $$ \widehat{\ourU}^{\alpha\beta} -\ourU^{\alpha\beta} =
-\frac{3\sqrt{|\fb|}}{8\pi} \zn_{\gamma} \left[ X^{[\gamma}
\zn^{\alpha}\zeta^{\beta]} + \zeta^{[\gamma}
\zn^{\alpha}X^{\beta]} \right] + \delta \ourU^{\alpha\beta}\;.
$$
(It would suffice that $\zRm_{\alpha\beta} = 2\Lambda
b_{\alpha\beta}/(n-1) +O(r^{-n/2})$ for the argument to go
through.) The first term on the right hand side above integrates
out to zero on $(n-1)$ dimensional boundaryless compact
submanifolds, and the remainder is order $o(r^{-n})$, so that
\begin{equation}
  \label{eq:*}
  \int_{ r =  R}  \hat \ourU^{\alpha\beta}dS_{\alpha\beta} =  \int_{ r
    =  R}   \ourU^{\alpha\beta}dS_{\alpha\beta} + o(1)\;,
\end{equation}
with $o(1)$ tending to zero as $r$ goes to infinity. We also have
\begin{equation}
  \label{eq*}
  \int_{ r =  R}  \hat \ourU^{\alpha\beta}dS_{\alpha\beta} =  \int_{
    \hat r
    =  \hat R}   \hat \ourU^{\alpha\beta}dS_{\alpha\beta}
  +2 \int_{V_{R,\hat R}}   \znabla_\alpha \hat
  \ourU^{\alpha\beta}dS_{\beta} \;,
\end{equation}
where $V_{R,\hat R}$ is a set the boundary of which is the union
of the coordinates sets $\{ r = R\}$ and $\{\hat r = \hat R\}$.
Conditions \eq{c1} guarantee that the volume integral in \Eq{eq*}
tends to zero when both $R$ and $\hat R$ tend to infinity, which
together with \eq{eq:*} establishes our claims.\qed

Let us finally show that the integrals \eq{toto} are {\em
covariant} under isometries of the background.  In what follows
$\hyp$ is an arbitrary hypersurface on which the charge integrals
\eq{toto} converge:
\begin{Lemma}
  \label{L1}
  Let $\Phi:\cM\to\cM$ be an isometric diffeomorphism of $(\cM,b)$
  such that $\Phi(\hyp)=\hyp.$ Then
  \begin{equation}
    \label{eq:l1}
    m(\hyp,\Phi^*g,b,(\Phi_*)^{-1}X) =  m(\hyp,g,b,X)\;.
  \end{equation}
\end{Lemma}

\proof Formula \eq{toto} for mass is invariant under
diffeomorphisms, hence
\begin{eqnarray*}
   m(\Phi(\hyp),g,b,\Phi_* Y) = m(\hyp, \Phi^*g,\Phi^*b,Y)\;,
\end{eqnarray*}
and the result follows from $\Phi(\hyp)=\hyp$, $\Phi^*b=b$.\qed

\section{Asymptotic isometries - the Riemannian problem}
\label{Sai}

Throughout this section, \emph{in contradistinction with the
remainder of this paper}, $g$ will denote the Riemannian metric
induced by the space-time metric on $\hyp$.  Similarly the letter
$b$ will denote the associated Riemannian background
metric\footnote{One could also consider background metrics of the
form $b= a^2(\bar r)d\bar r^2 +q^2(\bar r) h$; however, if $q$ is
sufficiently differentiable, then $b$ can be brought to the form
\eq{eq:b00} in the asymptotic region by a change of coordinates $r
= q(\bar r)$ provided that $dq/d\bar r$ has no zeros for large
$\bar r$'s.} of the form
\begin{equation}
  \label{eq:b00}
b=  a^2(r)dr^2 +r^2 h\;, \qquad h=h_{AB}(v^C)dv^A dv^B\;,
\end{equation}
with the indices $A$ running from $2$ to $n$. We assume that
$r\in[R,\infty)$ for some $R$, while the coordinates $v^A$ are
local coordinates on some compact $n-1$ dimensional manifold
$\mn$. Unless explicitly stated otherwise, we use the symbol
$\cO(r^\beta)$ to denote either $O(r^\beta)$ {\em throughout} this
section, {\em or} $o(r^\beta)$ {\em throughout} this section. We
shall mainly be interested in background metrics for which
\begin{deqarr}
& ra(r) =  \ell  + \newa(r)\;,\quad \newa = O(r^{-m_1})\;, \quad
\R\ni m_1>0\;, &
 \label{eq:b0-1}
\\ &  \newa'(r)= O(r^{-1-m_1})\;. &
\label{equicond} \arrlabel{eq:b0}\end{deqarr} for some constants
$m_1,\ell>0$.\footnote{ A typical example is
$a(r) = \frac \ell r\left(1 + \sum_{i=m_1}^{m_2}
  {a_i\over r^i} +\cO(r^{-\alpha})\right)
$ 
for some constants $m_1\ge 1$, $m_2\ge m_1$, $a_i$ and
$\alpha$.\label{aexp}  } When the vacuum Einstein equations
\eq{Ee} are satisfied by the metric \eq{eq:b00} we
have$^{\mathrm{\ref{footWoolgar}}}$ $a(r)=1/\sqrt{r^2/\ell^2+k}$,
where $k$ is a constant, which can be written in the form
\eq{eq:b0} with $m_1=2$, as well as in the form of
footnote~\ref{aexp} (with $m_2$ of that footnote as large as
desired). However, the hypothesis that the vacuum Einstein
equations hold plays no role in this section, therefore in
\eq{eq:b0} any $m_1>0$ will be allowed.
Let us mention that \eq{equicond} is equivalent to the  condition
\begin{equation}\newa(\hr) - \newa(r) = O(r^{-1-m_1})(\hr -r )
 \label{eq:b0-2m}
\end{equation}
 for $r$ large enough, and for
 $|\hr-r|\le r/2$. Indeed, under \eq{equicond} we have $$ \newa(\hr) -
 \newa(r) = \left(\int_0^1 \newa'(t\hr + (1-t)r)dt\right)(\hr -r)\;,$$
 and \Eq{eq:b0-2m} follows. The implication the other way round is
 straightforward using the fact that $\newa$ is smooth (recall that
local smoothness of the
 metric is assumed throughout). Condition~\eq{eq:b0-2m}  is actually
the one which is needed in the arguments below.

Let $\theta^i$, $i=1,\ldots,n$ be an orthonormal coframe for $b$,
with $\theta^1= a(r)dr$, and let $e_i$ be the dual frame; we
denote by
$$\zo^{i}{}_{jk}\equiv \theta^i(\zn_{e_k}e_j)$$ the associated
connection coefficients, where $\zn$ is the Levi-Civita connection
of $b$.  One easily finds
\begin{equation}
  \label{eq:b3}
  \zo^A{}_{1B}=  \frac 1 {ra(r)} \delta ^A_B= -\zo^1{}_{AB}\;.
\end{equation}
If we denote by \begin{equation}
  \label{eq:b2}\alpha^A\equiv \alpha(v^C)^A{}_{B} dv^B\end{equation}
an orthonormal frame for the metric $h$, and by $\beta^A{}_{BC}$
the associated Levi-Civita connection coefficients, then
\begin{equation}
  \label{eq:b4}
  \zo^A{}_{BC}= \frac 1r \beta^A{}_{BC}\;.
\end{equation}
All connection coefficients other than those in \eq{eq:b3} or
\eq{eq:b4}vanish.

\begin{Lemma}
  \label{L2} Let  $\theta^i$ be an orthonormal
  coframe for the metric $b$ as in \eq{eq:b00}, let
  $g=g_{ij}\theta^i\otimes \theta^j$, and suppose that
  $$g_{ij}\to_{r\to\infty} \delta^i_j\;.  $$ Denote by $\sR$,
  respectively $\sRz$, the $g$-geodesic distance along $\hyp$,
  respectively the $b$-geodesic distance along $\hyp$, from the set
  $\{r=R\}$. There exists a function $C(R)\ge 1$ satisfying
  $\lim_{R\to\infty}C(R)=1$ such that \begin{equation} \label{eq:l2}
  C(R)^{-1} \sRz \le \sR \le C(R) \sRz\;.  \end{equation}
\end{Lemma}

\medskip

\proof For $s\in[R,r]$ let $\gamma(s)=(s,v^A)$, then
\begin{eqnarray}
  \sR(r,v^A) & \le & \int_R^r \sqrt{g(\dot \gamma,\dot \gamma)(s,v^A)}\ds
\nn\\ & = & \int_R^r \sqrt{g_{ij}\theta^i(\dot
\gamma)\theta^j(\dot \gamma)(s,v^A)}\ds \nn\\ & = & \int_R^r
\sqrt{g_{11}(s,v^A)} a(s) \ds\label{fsr1}
\\\nn & \le & (1+o(1))\sRz(r,v^A)\;.
\end{eqnarray}
To obtain the reverse inequality, we note that for points
$(r,v^A)$ such that $r\ge R$ it holds that
$$
\forall X\quad g(X,X) \ge (1+o(1)) b(X,X)\;,$$ with $o(1)$ going
to zero as $R\to \infty$, hence for every curve $\gamma$ from
$\{r=R\}$ to $(r,v^A)$ we have
$$\int_\gamma \sqrt{g(\dot \gamma,\dot \gamma)(s)}\ds\ge
(1+o(1))\int_\gamma \sqrt{b(\dot \gamma,\dot \gamma)(s)}\ds\;,
$$
therefore
\begin{eqnarray}
\sR&=&\inf_\gamma \int_\gamma \sqrt{g(\dot \gamma,\dot
\gamma)(s)}\ds \nn\\ &\ge& (1+o(1)) \inf_\gamma \int_\gamma
\sqrt{b(\dot \gamma,\dot
  \gamma)(s)}\ds
\nn\\ &= & (1+o(1))\sRz\;. \label{fsr2}
\end{eqnarray} \qed

 The proof of Lemma~\ref{L2}
uses only the product structure of $b$, and does not require
Equation~\eq{eq:b0} to hold. If, however, that last equation
holds, then clearly
$$\sRz(r,v^A)= \int_R^r a(s)ds =\ell \ln (r/R) +O(R^{-m_1})\approx \ell \ln (r/R) $$ for large
$r$, and \eq{eq:l2} implies that for all $\epsilon>0$ there exists
$R_\epsilon\ge R$ and a constant $\hat C(\epsilon)$ such that for
all $r\ge R_\epsilon $ we have
\begin{equation} \label{eq:l2aold} \hat C(\epsilon)^{-1} r^{1-\epsilon}
\le \exp{(\sR/\ell)} \le \hat C(\epsilon) r^{1+\epsilon}\;.
\end{equation}
We will need a sharper version of this:

\begin{Lemma}
  \label{L2.1} Under the hypotheses of Lemma~\ref{L2}, suppose further that
  \Eq{eq:b0} holds, and that there exists a constant $\alpha>0$ such that
  $$g_{ij}-\delta^i_j=\cO(r^{-\alpha})\;.  $$ Then
  for $r\ge \max[R,1]$ we have
\begin{equation} \label{eq:l2a0} \exp{(\sR/\ell)} = r/R + O(R^{-m_1}) +
\cO(R^{-\alpha})\;,
\end{equation}
in particular
\begin{equation} \label{eq:l2a1}  r/C'
\le \exp{(\sR/\ell)} \le  C' r\;.
\end{equation}
\end{Lemma}

\proof
Here and elsewhere in this paper the letter $C$ denotes a constant
which may vary from line to line; if $\cO=o$ the constants in the
current proof can be chosen as small as desired by choosing $R$
large enough. In \Eq{fsr1} we can estimate $\sqrt{g_{11}}$ by $1+C
s^{-\alpha}$, obtaining thus
\begin{eqnarray*}
\sR(r,v^A) &\le &\sRz(r,v^A)+ CR^{-\alpha}
\\ &=& \ell \ln (r/R) + O(R^{-m_1}) + CR^{-\alpha}\;.
\end{eqnarray*}
Similarly, \Eq{fsr2} is rewritten as
\begin{eqnarray}
\sR&=&\inf_\gamma \int_\gamma \sqrt{g(\dot \gamma,\dot
\gamma)(s)}\ds \nn\\ &\ge&\inf_\gamma
\int_\gamma(1-Cs^{-\alpha})\sqrt{b(\dot \gamma,\dot \gamma)(s)}\ds
\;. \label{fsr3}
\end{eqnarray}
The last term in \Eq{fsr3} is the distance from the set $\{r=R\}$
in the metric
$$ (1-Cr^{-\alpha})^2\left(a^2(r)dr^2 +r^2 h\right)\;,$$
which equals
$$
\int_R^r (1-Cs^{-\alpha})a(s)\ds= \ell \ln(r/R) + O(R^{-m_1}) -
O(R^{-\alpha})\;,
$$
and our claims immediately follow. \qed

 The key result in our work is the following:
\begin{Theorem}[Asymptotic symmetries] \label{T1} Let $(r,v^A)$ coordinatize
  $\Omega\subset \hyp$ so that $\Omega
  \approx\{r\in[R,\infty)\}\times\mn$, and let $(\hr,\hv^A)$ be
  another set of coordinates on $\Omega$ so that
  $\Omega\approx\{(\hv^A)\in \hM ,\hr\in [\hat R(\hv^A),\infty)\}$ for
  some continuous function $\hat R$. We further assume that $v^A$ and
  $\hv^A$ are consistently oriented, and that $(r,v^A)$ and
  $(\hr,\hv^A)$ are also consistently oriented.  Let $b$, $\theta^i$,
  $e^i$, \emph{etc.}, be as at the beginning of this section, with $a$
  of the form
\eq{eq:b0}, and let $\hb$, $\hth^i$,
  $\he_i$, \emph{etc.} be the hatted equivalents thereof, so
  that $$\hb =
  a^2(\hr)d\hr^2 +\hr^2 \hh= \sum_i \hth^i\otimes\hth^i\;,
  $$
for some Riemannian metric $\hh$ on $\hM $. Suppose that there
exists $\alpha
>0$ such that
\begin{eqnarray}
  \label{eq:hd}
& g(e_i,e_j)-\delta_i^j = \cO(r^{-\alpha})\;,\quad e_k(g(e_i,e_j))
= \cO(r^{-\alpha})\;, & \\ &
  g(\he_i,\he_j)-\delta_i^j = \cO(\hr^{-\alpha})\;,\quad
\he_k(g(\he_i,\he_j)) = \cO(\hr^{-\alpha})\;.
\end{eqnarray}
Then:
\begin{enumerate}
\item There exists a $C^{\infty}$ map
$\Psi:\mn\to\hM $ satisfying
$$\Psi^*\hh = e^{-2\psi} h$$
for some $C^\infty$ function $\psi:\mn\to\R$, and\ptc{ would make
sense to add the second derivative equations here as well, and
give a coordinate counterpart of those second derivative equations
}
\begin{eqnarray}
  \label{eq:t1}& \hr = e^{\psi} r + O(r^{1-\beta})\;, \quad
  e_i (\hr) = e_i(e^{\psi}r)  +
O(r^{1-\beta})\;, & \\ & \hv^A = \psi^A(v^B)+ O(r^{-2})\;, \quad
e_i(\hat{v}^A) = e_i(\psi^A(v^B)) + O(r^{-2}) \;,  \label{eq:t2}
 &\end{eqnarray} in local coordinates with $\Psi =
(\psi^A)$, with $\beta=\min{(m_1,\alpha,2)}$.
\item If
$\Psi$ is the identity and if $\psi=0$, then for $\alpha >1$ we
further have
\begin{eqnarray}
  \label{t1}& \hr =  r + \cO(r^{1-\alpha})\;, \quad
  e_i (\hr) = e_i(r)  +
\cO(r^{1-\alpha})\;, & \\ & \hv^A = v^A+ \cO(r^{-\alpha-1})\;,
\quad e_i(\hat{v}^A) = e_i(v^A) + \cO(r^{-\alpha-1}) \;,
\label{t2}&
\\ &e_i\left(e_j(\hat{r}-r)\right) =
\cO(r^{1-\alpha})\;, \quad e_i\left(e_j(\hat{v}^A-v^A)\right) =
\cO(r^{-\alpha-1}) \;.  \label{t2a}&\end{eqnarray}
\end{enumerate}
\end{Theorem}

 \remarks
 1. For reference we note the partial derivatives estimates
 \begin{eqnarray}
  \label{eq:t1+1}& \displaystyle\frac{\partial \hr}{\partial r} = e^{\psi} + O(r^{-\beta})\;,
  \quad
   \displaystyle\frac{\partial \hr}{\partial v^A} =
   \frac{\partial e^{\psi}}{\partial v^A}r+ O(r^{1-\beta})\;,
   & \\ & \displaystyle\frac{\partial\hv^A}{\partial r} =  O(r^{-3})\;, \quad
\frac{\partial\hat{v}^A}{\partial v^B} =
\frac{\partial\psi^A}{\partial v^B}+ O(r^{-\beta}) \;,
\label{eq:t2+1}
 &\end{eqnarray}
with the second estimate in \eq{eq:t2+1} being somewhat stronger
than its counterpart in \eq{eq:t2}.

\noindent 2. It should be noted
 that in point 1. above we do not assume that $\hM =M$
and $\hh=h_{AB}(\hat v^C)d\hat v^Ad\hat v^B$; this fact plays a
role in~\cite{ChHerzlich}. The arguments in that last reference
show that $\Psi$ is a diffeomorphism, in particular $\hM$ is
necessarily diffeomorphic to $M$. If $\hh=h_{AB}(\hat v^C)d\hat
v^Ad\hat v^B$ and $\Psi$ is the identity, then clearly $\psi=0$
follows.

\medskip

\noindent 3. We stress that we do not assume $M$ or $\hM $ to be
parallelizable, thus \Eqs{eq:t1}{t2a} have to be understood in the
sense of finite coverings of $M$ and $\hM $, with corresponding
frames, on which the claimed estimates hold. However, if $M$ or
$\hM$ are parallelizable, then the estimates are global.
\medskip

\proof Let $\ncO_p$ be a conditionally compact subset of an open
domain of a local coordinate system $(v^A)$ around $p=(v_0^A)\in
M$, and let, on a neighbourhood of $\overline{\ncO_p}$, $\beta_A$
be  a $h$-orthonormal frame dual to $\alpha^A$. We note that the
connection coefficients $\beta^A{}_{BC}$ are uniformly bounded on
$\ncO_p$. Consider the radial ray $$\gamma_p(r):=(r,v_0^A)\;,$$
which, in hatted coordinates, can be written as
\be\label{gamhat}[R,\infty)\ni r\to
\gamma_p(r)=\left(\hr(r,v_0^A),\pgp(r):=\left(\hv^A(r,v_0^A)\right)\right)\in
[\hat R,\infty)\times \hM  \subset \hyp\;;\ee here and in what
follows we identify $[R,\infty)\times \ncO_p$ with the
corresponding subset of $\hyp$, similarly for sets of the form
$[\hat R,\infty)\times \widehat\mcU$, where $\widehat\mcU\subset
\hM $. It should be clear from \eq{gamhat} that the operation
``$\gamma_p\to\pgp$" above is a coordinate projection which
consists in forgetting the $\hr$ coordinate in a coordinate system
$(\hr,\hv^A)$. We have
\begin{eqnarray*}
\hh_{AB}\frac{\partial \hv^A}{\partial r}\frac{\partial
\hv^B}{\partial r} & \le & \frac 1 {\hr^2}\left(\hr^2
\hh_{AB}\frac{\partial \hv^A}{\partial r}\frac{\partial
\hv^B}{\partial r} + a^2(\hr)\left(\frac{ \partial \hr}{\partial
r}\right)^2\right)
\\ &= &\frac 1 {\hr^2} \hat b(\frac{ \partial }{\partial
r},\frac{ \partial }{\partial r}) \ \le \ C\frac 1 {\hr^2}
g(\frac{
\partial }{\partial r},\frac{ \partial }{\partial r})
\\ &\le  &C^2\frac 1 {\hr^2} b(\frac{ \partial }{\partial
r},\frac{ \partial }{\partial r})\ =\ C^2\frac {a^2(r)} {\hr^2}\;.
\end{eqnarray*}
Let $d_{\hh}$ denote the $\hh$ distance on $\hM $, for $r_2\ge
r_1$ we thus obtain
\begin{eqnarray}\label{hatdist}
d_{\hh}\left(\pgp(r_1),\pgp(r_2)\right) & \le & \int _{r_1}^{r_2}
\sqrt{\hh_{AB}\frac{\partial \hv^A}{\partial r}\frac{\partial
\hv^B}{\partial r}(s)}\; ds
\\ \nn&\le &C\int _{r_1}^{r_2}\frac {a(s)} {\hr(s,v_o^A)}\; ds
\\ \nn &\le& C^2\int _{r_1}^{r_2}\frac {1} {s^{2}}\; ds = C^2\left(
\frac {1} {r_1}-\frac {1} {r_2}\right)\;,
\end{eqnarray} and Lemma~\ref{L2.1} has been used.
 It then easily follows that
$$\hat p:= \lim_{r\to\infty}\pgp(r)$$ exists, with
\be\label{*} d_{\hh}\left(\pgp(r),\hat p\right)\le \frac {C^2}
{r}\;.\ee Let $\hOp$ be a conditionally compact subset of a domain
of local coordinates $\hv^A$ around $\hat p$, \Eq{*} shows that
$\gamma_p$ enters and remains in $[\hat R,\infty)\times \hOp$ for
$r$ large enough. In what follows only such $r$'s will be
considered.

Consider, now, a point $q\in\ncO_p$; we wish to show that the
corresponding ray $\gamma_q$ will stay within $[\hat
R,\infty)\times \cOp$ if $q$ is close enough to $p$. In order to
do that, consider an $h$-geodesic segment $\gamma \subset M$
parameterized by proper length such that $\gamma(0)=p$ and $
\gamma(d_h(p,q))=q$.
Expressing the path $$s\to\Gamma(s):=(r,\gamma(s))\in
[R,\infty)\times\ncO_p\subset \hyp$$ in terms of the barred
coordinates we have
\begin{eqnarray*}
\hh_{AB}\frac{d\hv^A}{ds}\frac{d \hv^B}{ds} & \le & \frac 1
{\hr^2}\left(\hr^2 \hh_{AB}\frac{d\hv^A}{ds}\frac{d \hv^B}{ds} +
a^2(\hr)\left(\frac{ d \hr}{ds}\right)^2\right)
\\ &= &\frac 1 {\hr^2} \hat b(\frac{ d\Gamma }{ds},\frac{d\Gamma}{ds})
\ \le \ C\; \frac 1 {\hr^2}  b(\frac{ d\Gamma
}{ds},\frac{d\Gamma}{ds})
\\ & = &C\; \frac {r^2} {\hr^2}\ \le\  C^2\;,
\end{eqnarray*}
An estimation as in \Eq{hatdist} gives
\begin{eqnarray*}
d_{\hh}\left(\pgp(r),\pgq(r)\right) & \le & C\;
d_h(p,q)\;.\end{eqnarray*} Passing to a subset of $\ncO_p$ if
necessary we thus obtain that for all $q\in \ncO_p$ the rays
$\gamma_q$ enter and remain in $[\hat R,\infty)\times \cOp$, for
$r\ge R_p$ for some $R_p$.

Let, on an open neighbourhood of $\overline{\cOp}$, $\hat\alpha^A$
be a $\hh$-ON frame with uniformly bounded connection coefficients
$\hat\beta^A{}_{BC}$, and let $\hat\beta_A$ be a $\hh$-orthonormal
frame dual to $\hat\alpha^A$. Equations \eq{eq:b3} and \eq{eq:b4}
show then that all the $\zo^i{}_{kj}$'s and $\hzo^i{}_{kj}$'s ---
the connection coefficients of $b$ and of $\hb$
--- are uniformly bounded along the rays $\gamma_q$, $q\in\ncO_p$; the reader will note that the same
will be true for the constants controlling various error terms
$\cO(r^{\cdot})$ in the calculations below. The idea of the
argument below is then to derive the desired estimates along the
$\gamma_q$'s, $q\in\ncO_p$; covering the compact manifold $\mn$ by
a finite number of coordinate patches $\ncO_{p_i}$,
$i=1,\ldots,I$, will establish our claims.

Let $f_i$, respectively $\hf_i$, be a $g$-orthonormal frame
obtained by a Gram-Schmidt orthonormalisation procedure using
$\{e_i\}_{i=1}^n$, respectively $\{\he_i\}_{i=1}^n$. The explicit
form of the $f_i$'s and $\hf_i$'s in terms of the $e_i$'s and
$\he_i$'s shows that
\begin{eqnarray}
&   f_i = e_i +\delta f_i\;,\quad \delta f_i = \delta f_i{^j}
e_j\;, \quad \delta f_i{^j} = \cO(r^{-\alpha})\;,\quad e_k(\delta
f_i{^j}) = \cO(r^{-\alpha}) \;, & \nn\\ && \label{eq:B1a}\\
& \hf_i = \he_i +\delta \hf_i\;,\quad \delta \hf_i = \delta
\hf_i{^j} \he_j\;, \quad \delta \hf_i{^j} =
\cO(\hr^{-\alpha})\;,\quad \he_k(\delta \hf_i{^j}) =
\cO(\hr^{-\alpha}) \;. \nn \\ && \label{eq:B1b}
\end{eqnarray}
By construction we actually have
 \bel{Gseq} f_1= \left(1+
\cO(r^{-\alpha})\right) e_1 =  \left(1+ \cO(r^{-\alpha})
+O(r^{-m_1} )\right)\frac r \ell \partial_r \;. \ee The uniform
boundedness of the $\zo^i{}_{kj}$'s further shows that
\be\label{eq:B1c} \zn_{e_i}\delta f_j = \cO(r^{-\alpha})\;,\ee
similarly for the $\hb$-covariant derivatives of the $\delta
\hf_j$'s with respect to the $\he_i$'s. Recall that
\be\label{b0}\zo^i{}_{kj}=\frac 12 \left\{\theta^j([e_i,e_k]) -
  \theta^i([e_k,e_j])-\theta^k([e_j,e_i]) \right\}\;;\ee
Equation~\eq{b0} together with its $g$-equivalent and
\eq{eq:B1a}-\eq{eq:B1c} imply \be\label{eq:B1d}\omega^i{}_{jk} =
\zo^i{}_{jk} + \cO(r^{-\alpha})\;,\ee similarly
\be\label{eq:B1e}\home^i{}_{jk} \equiv
\hat\phi^i(\nabla_{\hf_k}\hf_j)= \hzo^i{}_{jk} +
\cO(\hr^{-\alpha})\;.\ee
We use the symbols $\phi^i$ and $\hat\phi^i$ to denote coframes
dual to $f_i$ and $\hf_i$. Now, both the $f_i$'s and $\hf_i$'s are
orthonormal frames for $g$, hence there exists a field of rotation
matrices $\Lambda=(\Lambda_i{}^j) \in O(n)$ such that \be
\label{B4a+}\hf_i = \Lambda_i{}^j f_j\;.\ee
 We recall that for rotation matrices we
have\footnote{We use the convention summation throughout, so that
repeated indices in different positions have to be summed over. We
will explicitly indicate the summation only in those equations in
which we need to sum over repeated indices which are both
subscripts or both superscripts.}
\begin{equation}
  \label{eq:B4a}\sum_k \Lambda_i{}^k\Lambda_j{}^k = \delta^i_{j}\;,\ee
in particular
$$(\Lambda^{-1})_i{}^j = \Lambda _j {}^i\;,$$
so that $\hphi^i= \Lambda _j {}^i\phi^j$, and $\phi^j = \sum_i
\Lambda_j{}^i\hphi^i$. Further, from  \Eq{eq:B4a} we obtain
\begin{equation}
  \label{eq:B4} \sum_k \Lambda_i{}^k\Lambda_i{}^k=1\quad
  \Longrightarrow \quad \forall\; i,j\ \; |\Lambda_i{}^j| \le 1\;.
\end{equation}
From the definition of the connection coefficients we have
\begin{eqnarray*}
  \hat{\omega}^\ell{}_{ji} & = & \langle\hat \phi^\ell,\nabla_{\hf_i}
  \hf_j\rangle
\\  & = & \langle\Lambda_k{}^\ell \phi^k,\nabla_{\Lambda_i{}^m f_m}
  \left(\Lambda_j{}^n f_n\right)\rangle
\\  & = & \Lambda_k{}^\ell\Lambda_i{}^m \langle \phi^k,{f_m}(
  \Lambda_j{}^n) f_n+ \Lambda_j{}^n\nabla_{f_m}
  f_n\rangle \;,
\end{eqnarray*}
leading to the well known transformation law
$$\hat{\omega}^l{}_{ji} \Lambda_l{}^k = \Lambda_i{}^l
f_l(\Lambda_j{}^k) + \Lambda_j{}^l \Lambda_i{}^n
\omega^k{}_{ln}\;,
$$
which we intepret as an equation for the $\Lambda_j{}^k$'s:
\begin{equation}
  \label{eq:B5}
f_i(\Lambda_j{}^k) = (\Lambda^{-1})_i{}^l\hat{\omega}^n{}_{jl}
\Lambda_n{}^k - \Lambda_j{}^l \omega^k{}_{li}\;.
\end{equation}
From \Eqsone{eq:b0} and \eq{eq:B1a} we obtain
\begin{deqarr}
 \phi^m(\partial_r) &=& {a( r)} \left(\delta^m_1+
\cO(r^{-\alpha})\right)\label{eq:B6-}
\\ &=& \frac \ell r \left(\delta^m_1+
\cO(r^{-\beta})\right) \;,   \label{eq:B6}\arrlabel{B6}
\end{deqarr}
with $\beta=\min(\alpha,m_1)$, except if $\cO=o$ and $\alpha >m_1$
in which case either  $\beta$ should be taken to be any number
smaller than $m_1$, or $\cO$ should be understood as $O$.
Rescaling $r$ and the metric $g$ by a constant conformal factor
 we may without loss of
generality assume that $\ell=1$; similarly for $\hr$. \Eq{eq:l2a1}
together with Equations~\eq{eq:B1d}-\eq{B6} leads to
\begin{deqarr}
  \frac{\partial \Lambda_i{}^{j}}{\partial r} &=&
 {a( r)}
\left(
    \sum_k\hzo{}^\ell{} _{ik} \Lambda _\ell{}^j \Lambda _k {} ^1 +
    \cO(r^{-\alpha})\right) \label{eq:B7a-} \\ &=& \frac
  1 r  \left(
    \sum_k\hzo^\ell{} _{ik} \Lambda _\ell{}^j \Lambda _k {} ^1 +
    \cO(r^{-\beta})\right)\;,
  \label{eq:B7a}\arrlabel{B7a}
\end{deqarr}
in particular
\begin{deqarr}
   \frac{\partial \Lambda_1{}^{j}}{\partial r} &=& \frac
  {a( r)}{\hr a(\hr)} \left( \sum_A\Lambda _A{}^j \Lambda _A{} ^1+
  \cO(r^{-\alpha})\right) \label{eq:B8-}
\\&=& \frac
  1r \left( \sum_A\Lambda _A{}^j \Lambda _A{} ^1+
  \cO(r^{-\beta})\right)
\;.\label{eq:B8}\arrlabel{B8}
\end{deqarr}
Now, the transpose of a rotation matrix is again a rotation
matrix, therefore we also have
$$\sum_k\Lambda_k{}^i \Lambda_k {}^j = \delta^i_{j}\;,$$ which gives
\begin{equation}
  \label{eq:B8a}
  \sum_A\Lambda _A{}^1 \Lambda_A{} ^1 = 1 - (\Lambda_1{}^1)^2\;,
\end{equation}
and it follows that
\begin{deqarr}
   \frac{\partial \Lambda_1{}^{1}}{\partial r} &= & \frac
  {a( r)}{\hr a(\hr)} \left(1-
     (\Lambda _1{}^1)^2  +
    \cO(r^{-\alpha}) \right)
  \label{eq:B9-}
\\
 &= & \frac
  1r \left(1-
     (\Lambda _1{}^1)^2  \right)+
    \cO(r^{-\beta-1})
  \label{eq:B9}
\;.
  \arrlabel{B9}
\end{deqarr}
 We have the following:
\newcommand{\mysigma}{\sigma}
 \begin{Lemma}
   \label{L3} For all $\mysigma<\min(m_1,\alpha,2)$ we
   have
   \begin{eqnarray}
     \label{eq:B10}
     \Lambda _1{}^1 & = & 1+ \cO(r^{-\mysigma}) \;,
\\ r \frac{\partial \hr}{\partial r}& = & \hr +
\cO(r^{1-\mysigma})   \label{eq:B10a}\;.
   \end{eqnarray}
 \end{Lemma}
\proof Let $\chi$ denote the $ \cO(r^{-\beta-1}) $ term in
Equation~\eq{eq:B9}, set $g:=\Lambda _1{}^1$, and denote by
$\phi(r,v^A) = \int_r^\infty \chi (s,v^A)\ds= \cO(r^{-\beta})$;
Equation~\eq{eq:B9} shows that
$$\frac{\partial (g-\phi)}{\partial r} = \frac{1-g^2}{r}\ge 0\;.$$
It follows that $g-\phi$ is non-decreasing, and therefore has a
limit as $r$ goes to infinity; since $\phi$ tends to zero in this
limit we conclude that
$$g_\infty\equiv \lim_{r\to\infty}g$$
exists. Equation~\eq{eq:B9} shows that
\begin{equation}
  \label{eq:last}
  \lim_{r\to\infty} r {\partial g \over \partial r} =
1-g_\infty^2\;.
\end{equation}
Now Equation~\eq{eq:B4} gives $|g|\le 1$, while
Equation~\eq{eq:last} implies a logarithmic divergence of $g$
unless $g_\infty=\pm 1$; thus $g_\infty=\pm 1$. Define $f\ge 0$ by
the equation
$$g= g_\infty(1-f)\;,$$
then $f\to_{r\to\infty} 0$ and we have
$${\partial f \over \partial r} = -{g_\infty f(2-f) \over r} -
g_\infty\chi\;.$$ Suppose, first, that $g_\infty =1$; since
$f\to_{r\to\infty}=0$ it follows that for every $\delta>0$ there
exists $r_\delta$ such that $f\le \delta$ for $r\ge r_\delta$; for
such $r$ we then obtain
\begin{equation}
  \label{eq:B11}
  {\partial f \over \partial r} \le -{ f(2-\delta) \over r} -
\chi\;,
\end{equation}
and by integration
\begin{eqnarray*}
  r^{2-\delta} f(r,\cdot) - r_\delta^{2-\delta} f(r_\delta,\cdot) \le
  -\int_{r_\delta}^r \chi(s,\cdot)s^{2-\delta}\ds =
  \cO(r^{2-\delta-\beta}) + \cO(r_\delta^{2-\delta-\beta})\;,
\end{eqnarray*}
so that
\begin{equation}
  \label{eq:B12a}
  f(r) = \cO(r^{\delta-2}) + \cO(r^{ -\beta})\;.
\end{equation}
Choosing $\delta$ appropriately we obtain \eq{eq:B10} with any
$\mysigma<\min(m_1,\alpha,2)$, under the assumption that
$g_\infty\ne -1$. In the case $g_\infty= -1$ similar, but simpler,
manipulations lead again to Equation~\eq{eq:B12a} with $\delta=0$.
From Equation~\eq{eq:B8a} and from $ \Lambda_1{}^1 = g_\infty +
\cO(r^{-\mysigma})$ we obtain
\begin{equation}
  \label{eq:B14a}
  \Lambda_A{}^1 = \cO(r^{-\mysigma/2})\;.
\end{equation}
Equation~\eq{eq:B4a} similarly implies
\begin{equation}
  \label{eq:B14b}
  \Lambda_1{}^A = \cO(r^{-\mysigma/2})\;.
\end{equation}
Equation~\eq{eq:B7a-} gives
\begin{eqnarray*} 
  \frac{\partial \Lambda_A{}^{1}}{\partial r}
&  =& \frac 1 r \left(
    - \Lambda_1 {}^1\Lambda_A {}^1+
    \cO(r^{-\mysigma})\right) \;,
\end{eqnarray*}
and integration in $r$ together with \eq{eq:B14a} yields (assuming
without loss of generality that $\sigma\ne 1$)
\begin{equation}
  \label{eq:B15-}
  \Lambda_A{}^1 = O(r^{-1}) + \cO(r^{-\mysigma})\;.
\end{equation}
\ptc{unnecessary equation and redefinition of sigma removed}
Integrating in $r$ Equation~\eq{eq:B8} and using
\eq{eq:B14b}-\eq{eq:B15-} we similarly obtain\ptc{term added since
sigma's redefinition removed}
\begin{equation}
  \label{eq:B14}
  \Lambda_1{}^A = O(r^{-1})+\cO(r^{-\mysigma})\;.
\end{equation}
From the definition of the $\hf_i$'s and from \eq{eq:b0} (with
$\ell=1$) we have
$$\hf_1(\hr) =  \hr + \cO(\hr^{1-\beta})\;, \quad
\hf_A(\hr)=\cO(\hr^{1-\alpha})\;,$$ hence\ptc{$f_A(\hr)$ equation
commented out since not needed; still in the file}
\begin{eqnarray*}
  f_1(\hr) & =&  \Lambda_1{}^1 \hf_1(\hr) +  \Lambda_1{}^A \hf_A(\hr)
\\ & = & g_\infty  \hr + \cO(r^{1-\mysigma})\;,
\end{eqnarray*}
Inverting \Eq{Gseq} we have
\begin{eqnarray*}
\left(1+ O(r^{-m_1})\right) r {\partial
    \hr \over \partial r}
  & = &  e_1(\hr)
\\ & = &\left(1+\cO(r^{-\alpha})\right) f_1(\hr)
\\ & = & g_\infty \hr + \cO(r^{1-\mysigma})\;.
\end{eqnarray*}
We have finally obtained
\begin{equation}
  \label{eq:B16}
  r {\partial \hr \over \partial r } = g_\infty\hr
  +\cO(r^{1-\mysigma}) \;,
\end{equation}
which is compatible with the fact that the coordinate systems
$(r,v^A)$ and $(\hr,\hv^A)$, as well as  $(v^A)$ and $(\hv^A)$,
carry the same orientations if and only if $g_\infty=1$, and the
lemma is established. \qed

Returning to the proof of Theorem~\ref{T1}, it is useful to
introduce new coordinates $x$ and $\hx$ defined as
$$r=e^x\;, \qquad \hr = e^{\hx}\;.$$
\newcommand{\mysigmax}{\mysigma x}
In terms of those variables Equation~\eq{eq:B10a} can be rewritten
as \be\label{hxx}{\partial \hx \over \partial x } = 1 + \phi\;,
\qquad\phi =\cO(e^{-\mysigmax})\;,\ee for an appropriately defined
function $\phi$: More precisely, if we write \be\label{fnotation}
e_i=e_i{}^jf_j\;,\quad f_i= f_i{}^je_j\;, \ee with the obvious
hatted equivalents, then \be\label{phiequation} \phi:= \frac
{ra(r)}{\hr a(\hr)} \sum_{k,j} e_1{}^j \Lambda^j{}_k\hat
f_k{}^1-1\;.\ee
 Integration of \Eq{hxx} gives
\begin{eqnarray}
 \hx(x,v^A)& = & x-\hx(x_0,v^A) + \int_{x_0}  ^x \phi(s,v^A)\ds
\nonumber \\ & = &
 x-\hx(x_0,v^A) + \int_{x_0}  ^\infty  \phi(s,v^A)\ds- \int_{x}  ^\infty \phi(s,v^A)\ds
\nonumber \\ & =: & x+\psi(v^A) + \cO(e^{-\mysigmax})\;,
\label{eq:B16a}
\end{eqnarray}
which establishes the existence of a continuous function
$\psi:\mn\to\R$ such that
\begin{equation}
  \label{eq:B17}
  \hr(r,v^A) = e^{\psi(v^A)} r + \cO(r^{1-\mysigma})
\end{equation}
(continuity of $\psi$ follows from the Lebesgue (dominated)
theorem on continuity of integrals with parameters; the Lebesgue
theorem is used in a similar way without explicit reference at
several places below).

 Let us write
\Eq{eq:B9-}  as
\begin{eqnarray}
   \frac{\partial (\Lambda_1{}^{1}-1)}{\partial r} &= &
\chi_1(\Lambda_1{}^{1}-1) +\chi_2 \;,
  \label{eq:B9-a}
\end{eqnarray}
where
$$ \chi_1:= -(1+\Lambda_1{}^{1})\frac
  {a( r)}{\hr a(\hr)} =-{2\over r}+\cO(r^{-1-\sigma})\;,
\qquad \chi_2 = \cO(r^{-\alpha-1})\;;$$ we obtain
\begin{eqnarray}
   { \Lambda_1{}^{1}}{(r,v^A)} &= & 1 + \left(f_1{}^{1}(v^A)
+ \int_{r_0}^r
s^2\exp\left\{\int_s^{\infty}\left(\chi_1(v,v^A)+{2\over
v}\right)dv\right\}\chi_2(s,v^A)ds\right)\times
\nonumber \\
& & r^{-2}\exp\left\{-\int_{r}^\infty\left(\chi_1(v,v^A)+{2\over
v}\right)dv\right\} \;,
  \label{eq:B9-b}
\end{eqnarray}
with some continuous function $f_1{}^{1}:M\to \R$. In particular
\begin{eqnarray}
\Lambda_1{}^1 &= & 1 + O(r^{-2})+\cO(r^{-\alpha})+ \delta
_\alpha^2\cO(r^{-\alpha}\ln r)
\;. \label{eq:B7a--}
\end{eqnarray}
From \Eq{eq:B7a-} we obtain
\begin{eqnarray}
  \frac{\partial \Lambda_A{}^{1}}{\partial r} &=& \frac
  {a( r)}{\hr a(\hr)} \left(-\Lambda _1{}^1 \Lambda _A {} ^1 +
  O(r^{-2})+
    \cO(r^{-2\sigma})+\cO(r^{-\alpha})\right)\;, \label{eq:B7a-a}
\end{eqnarray}
which integrated in a manner similar to that for \Eq{eq:B9-a}
shows that there exists a continuous function $f_A{}^1: M\to \R$
such that\ptc{terms added since $\alpha \le 1$ allowed now; there
was an error here before anyway, so \eq{deltaeq} added }
\begin{deqarr}
\Lambda_A{}^1(r,v^B) &= &
\frac{f_A{}^1(v^B)}{r}+\cO(r^{-\alpha})+\delta_1^\alpha \cO(
r^{-\alpha}\ln r) + O(r^{-1-\delta}) \nn \\ &&
 \label{eq:B8--1}\\
& = & O(r^{-1})+\delta_1^\alpha \cO(r^{-\alpha}\ln
r)+\cO(r^{-\alpha})\; , \label{eq:B8--}
\end{deqarr}
with any $\delta$ satisfying\ptc{maybe sigma $\le$ and not $<$?}
\bel{deltaeq} \delta<\min(1,2\sigma-1)\;.\ee It is easily seen now
that the $r^{-2}\ln r$ terms which could potentially be present in
\Eq{eq:B7a--} cannot occur: clearly they are only relevant for
$\alpha = 2$; in that last case it immediately follows from
\Eqsone{eq:B8a} and \eq{eq:B8--1} that no such terms are allowed.
It follows that
\begin{eqnarray}
\Lambda_1{}^1 &= & 1 + O(r^{-2})+\cO(r^{-\alpha})\;. \label{B7a--}
\end{eqnarray}
Repeating the argument after \Eq{eq:B14} one is led to
\begin{deqarr}
{\partial \hr \over \partial r} & = & {\hr \over r} 
+
\cO(r^{-\alpha}) +O(r^{-m_1}) + O(r^{-2})\;, \label{eq:B18first}\\
\hr &=& e^{\psi(v^A)} r + \cO(r^{1-\alpha})+O(r^{1-m_1}) +
O(r^{-1})\;;
  \label{eq:B18}\arrlabel{eq:B18m}
\end{deqarr}
\eq{eq:B18} has been obtained from \eq{eq:B18first} by calculating
$\partial \hx / \partial x$ and integrating the resulting
equation, compare \Eqs{hxx}{eq:B16a}. Equations~\eq{eq:B7a-} and
\eq{eq:B18} yield\ptc{log term added cause of small alphas}
\begin{eqnarray*}
  {\partial \Lambda_A{}^B \over \partial r} &= & a(r) \left(
    \sum_k\hzome^\ell{}_{Ak} \Lambda_\ell{}^B\Lambda_k{}^1
    +\cO(r^{-\alpha})\right)
\\ &= & a(r) \left(
    \sum_C\hzome^\ell{}_{AC} \Lambda_\ell{}^B\Lambda_C{}^1
    +\cO(r^{-\alpha})\right)
\\ & = & \cO(r^{-\alpha-1})
+\delta_1^\alpha \cO(r^{-\alpha-1}\ln r) +O(r^{-3})\;,
\end{eqnarray*}
and by integration one finds that there exists a continuous matrix
valued function $R=(R_A{}^B):\mn\to O(n-1)$ such that
\begin{equation}
  \label{eq:B19}
  \Lambda_A{}^B(r,v^C)= R_A{}^B(v^C)+  \cO(r^{-\alpha})+
\delta_1^\alpha \cO(r^{-\alpha}\ln^2 r) +O(r^{-2})\;.
\end{equation}
Repeating the argument which led to \Eq{eq:B14} and using
\eq{eq:B19} one  finds now that there exists a continuous function
$f_1{}^A: M\to \R$ such that\ptc{there is a problem with the delta
terms which might need clarifying; there was a wrong  log term in
the previous equation}
\begin{eqnarray}
\Lambda_1{}^A (r,v^B)&= &
\frac{f_1{}^A(v^B)}{r}\left(1+O(r^{-m_1})\right) \nn \\&&
+\cO(r^{-\alpha})+ \delta _\alpha^1\cO(r^{-\alpha}\ln r)
+ \cO(r^{-2\sigma})\;, \label{eq:B8--ao}
\end{eqnarray}
compare \Eq{eq:B7a-a}; without loss of generality we have assumed
that $\sigma \ne 1$. The hatted equivalent of \eq{eq:b2},
$$\hat\alpha^A\equiv \hat \alpha(\hv^C)^A{}_{B} d\hv^B\;,$$ gives
\begin{eqnarray*}
  d\hv^A & = & \hbeta^A{}_B \hat \alpha ^B
=\frac 1 \hr \hbeta^A{}_B\hat \theta ^B
\\ & = & \frac 1 \hr \hbeta^A{}_B
\left((1 +\cO(r^{-\alpha}))\Lambda_C{}^B \theta^C + (1
  +\cO(r^{-\alpha}))\Lambda_1{}^B \theta^1\right)\;,
\end{eqnarray*}
where $\beta^A{}_B$ denotes the matrix inverse to $\alpha^A{}_B$,
while the symbol $\hbeta^A{}_B$ is used to denote the the matrix
inverse to $\hat \alpha^A{}_B$. This implies\begin{eqnarray}
  \label{eq:B20}
  {\partial \hv ^A \over \partial r } &= &O(r^{-3})\;,
\\ \label{eq:B21}
  {\partial \hv ^A \over \partial v^C } &= &\frac r {\hr}
\hbeta^A{}_B \Lambda_D{}^B \alpha ^D{_C}  +\cO(r^{-\alpha}) \;.
\end{eqnarray}
Integrating \eq{eq:B20} in $r$ one obtains that the limits
$$\psi^A\equiv \lim_{r\to\infty}\hv ^A$$
exist and are continuous functions, with \bel{eq:B21+}
\hv^A-\psi^A = O(r^{-2})\;.\ee Moreover, it follows from
\eq{eq:B21} that the limits as ${r}$ tends to infinity of $
{\partial \hv^A /
\partial v^B}$ exist and are continuous.\ptc{Argument rearranged,
since the previous one did not work for small alpha}
 Passing
to the limit $r\to\infty$ in Equation~\eq{eq:B21} one obtains
\bel{B22++}\Psi^*\hat \alpha^A= e^{-\psi} R^A{}_B \alpha ^B\;,\ee
hence
\begin{eqnarray*}
  \Psi^*\hat h &= &\Psi^*\sum_A\hat\alpha^A\otimes\hat\alpha^A
\\ & = & e^{-2\psi} \sum_AR^A{}_B R^A{}_C \alpha ^B\otimes\alpha ^C
\\ & = & e^{-2\psi} \sum_A\alpha^A\otimes\alpha^A = e^{-2\psi} h\;,
\end{eqnarray*}
where we have used the fact that $R=(R_A{}^B)$ is a rotation
matrix. It follows that the map $\Psi=(\psi^A)$ is a conformal
local diffeomorphism from $(M,h)$ to $(\hM,\hh)$. We can thus use
a deep result  of Lelong-Ferrand~\cite{Lelong-Ferrand} to conclude
that $\Psi$ is smooth, in particular so is $\psi$.
Equation~\eq{B22++} shows then smoothness of $R^A{}_B$. Further
\begin{eqnarray}
  \frac 1 r \beta^B{}_A {\partial \hr \over \partial v^B}
 & = & e_A(\hr) \nonumber \\ & = & (1
+\cO(r^{-\alpha}))\sum_B\Lambda_B{}^A \hf_B(\hr)
 + (1 +\cO(r^{-\alpha}))\Lambda_1{}^A \hf_1(\hr)
\nonumber\\ & = & O(1)+\cO(r^{1-\alpha})\;, \label{B22}
\end{eqnarray}
hence \bel{B22+} {\partial \hx \over \partial v^A} = \frac 1 \hr
{\partial \hr \over \partial v^A} = O(1)+\cO(r^{1-\alpha})\;. \ee
This, together with \Eqs{eq:B20}{eq:B21+}, establishes point 1 for
$0<\alpha\le 1$.

If $\alpha>1$, a closer inspection of the $O(1)$ terms in
\Eq{B22+}, making use of \Eq{eq:B8--ao}, shows that the limits
$\lim_{r\to\infty} {\partial \hx / \partial v^A}$ exist, and are
continuous functions of the $v^A$'s. Now, in the current range of
$\alpha$'s it is easy to  show that the function $\psi$ in
\eq{eq:B16a} is continuously differentiable without invoking the
Lelong-Ferrand theorem, as follows: Let $\phi$ be the function
appearing at the right-hand-side of \Eq{hxx}, from
\eq{phiequation} and from what has been said with a little work
one finds
$$\frac{\partial \phi}{\partial v^A} = \cO(e^{(1-\alpha)x}) + O(e^{-x})\;; $$
the differentiability of $\psi$ follows now from its definition
\eq{eq:B16a} and from Lebesgue's dominated theorem on
differentiability of integrals with parameters. The last estimate
together with \Eq{eq:B16a} also show that
\be\label{hxvder}\frac{\partial \hx}{\partial v^A} =
\frac{\partial\psi}{\partial v^A}+\cO(e^{(1-\alpha)x}) +
O(e^{-x})\;.\ee \ptc{end of new stuff; a spurious piece of
previous  argument commented out} and point 1 is established.


To establish point 2, suppose that $\Psi$ is the identity and that
$\psi=0$. The calculation in \Eq{B22} shows that
$$0 =\lim_{r\to\infty} {\partial \hx \over
\partial v^A} =
\lim_{r\to\infty} {1\over \hr}{\partial \hr \over \partial v^A} =
\lim_{r\to\infty} r\alpha^B{}_A e_B(\hr)
 =\lim_{r\to\infty} \alpha^B{}_A\Lambda_B{}^1\;,
$$
hence the function $f_A{}^1$ appearing in \Eq{eq:B8--} vanishes.
The identity
\begin{equation}
\Lambda_1{}^A\Lambda_1{}^1+\sum_B\Lambda_B{}^A\Lambda_B{}^1 =0
\label{B23-}\end{equation} shows that the function $f_1{}^A$ from
\Eq{eq:B8--ao} vanishes as well. If $1<\alpha<2$ we thus obtain
\begin{equation}
\Lambda_i{}^j =\delta_i^j + \cO(r^{-\alpha})\;. \label{B23}
\end{equation}
\ptc{From now on changes} For $\alpha\ge2$ a closer inspection of
\Eq{eq:B8-} is needed: \begin{eqnarray} \nn \frac{\partial
\Lambda_1{}^A}{\partial r} &=&  \frac 1 r \frac {ra(r)}{\hr a
(\hr)} \left( \sum_C\Lambda_C{}^A\Lambda_C{}^1 +
\cO(r^{-\alpha})\right)
\\
&=&  \frac 1 r \frac {ra(r)}{\hr a (\hr)} \left(- \Lambda_1{}^1
\Lambda_1{}^A + \cO(r^{-\alpha})\right) \;,
\label{B8-++}\end{eqnarray} where we have used \Eq{B23-}. \ptc{an
elegant but unneeded argument commented out}
Integrating this equation in a way somewhat similar to
\Eq{eq:B9-a} shows that
\begin{eqnarray}
\Lambda_1{}^A (r,v^B)= \cO(r^{-\alpha})+ \delta
_\alpha^2\cO(r^{-\alpha}\ln r)\;. \label{eq:B8--ao+}
\end{eqnarray}
If $\alpha=2$, suppose for the moment that there is no $\ln r$
term in \eq{eq:B8--ao+}; it then follows from \Eqsone{B7a--} and
\eq{B23-} that \Eq{B23} holds. On the other hand, for $\alpha>2$
\Eq{eq:B8a} forces the function $f_1{}^1$ from \Eq{eq:B9-b} to
vanish, which in turn implies that \Eq{B23} holds again. The
formula inverse to \Eq{B4a+} reads \be \label{B4b+}f_j =
\sum_i\Lambda_i{}^j \hf_i\;,\ee in particular
$$f_1(\hr) = \Lambda_1{}^1 \hf_1(\hr)+\sum_A\Lambda_A{}^1 \hf_A(\hr)\;,$$
which implies
\begin{equation}{\partial \hr \over \partial r } = {a(r) \over a(\hr) }
\left(1+\cO(r^{ -\alpha})\right)+\cO(r^{ -\alpha})\;. \label{B24}
\end{equation}
\Eq{eq:b0-2m} together with the identities
\begin{deqarr}
\arrlabel{B24+}
 & \displaystyle
{ra(r) \over \hr a(\hr)} = { ra(r) - \hr a(\hr)\over \hr a(\hr)}
+1 = 1 + O(r^{-m_1-1}) \delta r \;, & \\ & \displaystyle {a(r)
\over  a(\hr)} = { ra(r) \over \hr a(\hr)} \times {\hr \over r} =
{ ra(r) \over \hr a(\hr)}  \left(1+{\delta r\over r}\right) = 1 +
\left({1\over r} +O(r^{-m_1-1})\right) \delta r \;.\end{deqarr}
shows that \Eq{B24} can be rewritten as
\begin{equation}
{\partial \delta r \over \partial r} = \chi_3
 \delta r + \chi_4 \;,
\label{B25}
\end{equation}
with
$$\chi_3 = {1\over r} + O (r^{-m_1-1})\;, \qquad \chi_4 =
\cO(r^{-\alpha})\;.
$$
Hence, for $r_2>r$ we have
\begin{eqnarray*}
{\delta r(r_2) \over r_2} &=& \left( {\delta r (r)\over r} +
\int_r^{r_2} \exp\left\{-\int_t^\infty \left(\chi_3(s)-{1\over
s}\right)ds\right\} {\chi_4(t)\over t}dt\right) \times\\ &&
\exp\left\{\int_r^\infty \left(\chi_3(s)-{1\over
s}\right)ds\right\} \;.
\end{eqnarray*}
Passing with $r_2$ to infinity and using the fact that $\delta r =
o(r)$ shows that
\begin{eqnarray}
 \delta r &=& -r \int_r^{\infty}
\exp\left\{-\int_t^\infty \left(\chi_3(s)-{1\over
s}\right)ds\right\} {\chi_4(t)\over t}dt \nonumber \\ &=&
\cO(r^{1-\alpha})\;. \label{B27-}\end{eqnarray} From \Eq{B24} we
obtain
\begin{equation}{\partial \delta r \over \partial r } = \cO(r^{ -\alpha})\;.
\label{B27}
\end{equation}
\Eq{B4b+} implies
\begin{eqnarray}
e_j &=&\left(\delta_j^k +\cO(r^{ -\alpha})\right) f_k \nonumber \\
&=& \sum_i\left(\delta_j^k +\cO(r^{ -\alpha})\right) \Lambda_i{}^k
\left(\delta_i^k +\cO(r^{ -\alpha})\right)
 \he_k
\nonumber \\ &=& \he_j +\sum_i \cO(r^{ -\alpha}) \he_i \;.
\label{B4c+}\end{eqnarray} It follows that
\begin{deqarr}
e_j(\hr) &=& \delta_j^1\hr +\cO(r^{1 -\alpha})
 \nonumber\\ &=&
e_j(r)  +\cO(r^{1 -\alpha})\;,
 \\
e_j(\hv^A) &=& \he_j(\hv^A)
 +\cO(r^{-1 -\alpha})
\;. \label{B4d+}
\end{deqarr}
In particular
\begin{eqnarray}
&\displaystyle{\partial \hv^A \over \partial r}  =  a(r)
e_1(\hv^A) = \cO(r^{-2-\alpha})
\quad \Longrightarrow \quad
 \hv^A - v^A = \cO(r^{-1-\alpha})\;.
\label{B4e+}\end{eqnarray} \Eqsone{B27-} and \eq{B4e+} allow us to
rewrite \Eq{B4d+} as
\begin{eqnarray}
e_j(\hv^A) &=& e_j(v^A)
 +\cO(r^{-1 -\alpha})
\;. \label{B4f+}\end{eqnarray} \Eqs{eq:t1}{eq:t2} are thus
established. A straightforward analysis of the equations
\begin{eqnarray*}
\label{B4g+}e_k\left(f_j(\hr)\right) &=&
e_k\left(\sum_i\Lambda_i{}^j \hf_i(\hr)\right)
\;,\\
e_k\left(f_j(\hv^A)\right) &=& e_k\left(\sum_i\Lambda_i{}^j
\hf_i(\hv^A)\right) \;,
\end{eqnarray*}
leads to \Eq{t2a}, and the theorem is established for $\alpha\ne
2$, or for $\alpha=2$ provided that no log terms are present in
\eq{eq:B8--ao+}.\reword

Let us return to the case $\alpha=2$; then \eq{B23} holds with any
$\alpha<2$ and therefore the calculations that follow remain valid
with any $\alpha<2$. Further \eq{B23} holds with $i=j=1$ and
$\alpha=2$ by \Eq{B7a--}. \Eqsone{eq:B5} and \eq{B24+} give then
$$\frac{\partial(r\Lambda_A{}^1)}{\partial r} = \cO(r^{-2}) \quad
\Longrightarrow \quad \Lambda_A{}^1= \cO(r^{-2})\;,
$$
so that no $\log$ terms can occur in $\Lambda_A{}^1$. \Eq{B23-}
implies then
$$ \Lambda_1{}^A= \cO(r^{-2})\;,
$$
and \eq{eq:B5} establishes \eq{B23} with $\alpha=2$,\reword and
the theorem follows. \qed

 In the next section we shall need the following:
\begin{Corollary}
\label{C1} Let $\Psi(r,v^A)=(\hr(r,v^A),\hv^B(r,v^A))$
 be an isometry of the background metric $b$:
$$\Psi^*(a^2(r)dr^2 +r^2 h) =a^2(\hr)d\hr^2 +\hr^2 \hh\;, \qquad \hh=h_{AB}(\hv^C)d\hv^A
d\hv^B\;.$$ If there exists $\alpha >1$ such that the physical
metric $g$ satisfies
\begin{eqnarray}
  \label{eq:hd1}
& g(e_i,e_j)-\delta_i^j = \cO(r^{-\alpha})\;,\quad e_k(g(e_i,e_j))
= \cO(r^{-\alpha})\;, &
\end{eqnarray}
where $e_i$, $i=1,\ldots,n$ is the usual ON frame for the metric
$b$ as in \eq{frame}, then
\begin{eqnarray}
  \label{eq:hd2}
 &
  g(\he_i,\he_j)-\delta_i^j = \cO(\hr^{-\alpha})\;,\quad
\he_k(g(\he_i,\he_j)) = \cO(\hr^{-\alpha})\;,
\end{eqnarray}
where $\he_i$ is the corresponding hatted frame.
\end{Corollary}

\proof Applying point 1. of Theorem~\ref{T1} to $g=b$ we obtain
that
\begin{eqnarray}
  \label{eq:t1+}& \hr = e^{\psi} r + O(r^{1-\beta})\;,
\end{eqnarray}
with $\beta=\min{(m_1,2)}$. Since $\Psi$ is an isometry we have
$\he_i = \Lambda_i{}^je^j$ for some rotation matrix
$\Lambda_i{}^j$, which gives
$$ \he^{ij}= (g-b)(\htheta^i,\htheta^j)
=\Lambda_k{}^i\Lambda_\ell{}^j(g-b)(\theta^k,\theta^\ell)=
\cO(r^{-\alpha})=\cO(\hr^{-\alpha})\;.$$ Further,
$$e_r({\he^{ij}})=e_r\left( \Lambda_k{}^i\Lambda_\ell{}^je^{kl}\right)\;,$$
and --- since $e_i( \Lambda_k{}^j)=O(1)$ by \eq{eq:B5} --- the
result easily follows. \qed
\section{Global charges}\label{Sgc}
\newcommand{\Isot}{\text{Iso}^\uparrow(\hyp,b)}
\newcommand{\Iso}{\text{Iso}(\hyp,b)}
\newcommand{\Isoo}{\text{Iso}_0(\hyp,b)}
\newcommand{\cC}{\mycal C}
Let us give here a general prescription how to assign global
geometric invariants to hypersurfaces $\hyp$ in the class of
space-times with metrics asymptotic to backgrounds \eq{A1}:
consider such a background metric $b$ and consider a hypersurface
$\hyp$ given by the equation $\{t=0\}$ in the coordinate system of
\eq{A1}. Let $\cK$ denote the set of all Killing vector fields of
$b$; the hypersurface $\hyp$ singles out two subsets of $\cK$: a)
the set $\cKS$ of those Killing vector fields of $b$ which are
normal to $\hyp$, and the set $\cKt$ of all $b$-Killing vector
fields which are tangent to $\hyp$.  Consider any metric $g$ for
which the fall-off hypotheses of Theorem~\ref{T0} are met, with
$X\in\cKS$, or with $X\in\cKt$, or perhaps with all $X\in\cK$. (In
that theorem we have assumed that $b$ satisfies Equation~\eq{Ee},
but it would be sufficient that \eq{Ee} holds only up to terms
which decay sufficiently fast when $r$ tends to infinity, the same
for $g$.)  Let $\Isot$ be the group of all time-orientation
preserving isometries of $b$ which leave $\hyp$
invariant.\footnote{Some further invariants can sometimes be
obtained by considering the connected component of the identity of
$\Isot$, but this seems to require a case by case analysis, so
that no general discussion will be given here.} We shall suppose
further that the following condition holds:
\begin{eqnarray}\nonumber
  &  \text{\emph{for every orientation-preserving
   conformal  isometry $\Psi$ of $(M,h)$ there exists  }}  & \\
\nonumber &\text{\emph{$R_*>0$ and a $b$-isometric map $\Phi:
[R_*,\infty)\times M \to [R,\infty)\times M$, such that}} & \\
\label{H} & \lim_{r\to\infty} \Phi(r,\cdot) = \Psi(\cdot)\;.&
\end{eqnarray} It follows from
\cite[Vol.~II,Theorem~18.10.4]{Berger} and from what is said in
Appendix \ref{Aai1} that this condition holds for the
$(n+1)$-dimensional anti-de Sitter metrics, $n\ge 3$; the case
$n=2$ is handled by the discussion of toroidal $\mn$'s below.
Further, the above condition obviously holds for those metrics for
which every conformal isometry of $(M,h)$ is an isometry, as is
the case for the $(M,h)$'s considered in Appendix~\ref{Aai2}: the
desired extension $\Phi$ is
$$\Phi(r,v^A)=(r,\Psi(v^A))\;.$$
In fact, it is shown in \cite{ChHerzlich} that condition~\eq{H}
always holds when $a(r)=1/\sqrt{r^2+k}$ regardless of the metric
$h$.

Let $\cC$ denote the collection of positively oriented coordinate
systems $c=\left(\cO,(r,v^A)\right)$, where $\cO$ is the domain of
definition of the collection of functions $(r,v^A)$, with the
associated background metrics and orthonormal tetrads, for which
Equations~\eq{c1}, \eq{eq:deccond} and \eq{eq:deccond1} hold.  For
each such coordinate system $c$ we can calculate the set of
integrals \eq{toto}, where $X$ runs over $\cKS$, or over $\cKt$,
or over $\cK$, whichever appropriate. By Theorem~\ref{T1} every
two coordinate system $c_1,c_2$ in $\cC$ differ by a coordinate
transformation, say $\Upsilon$, the $M$-part of which asymptotes
to an orientation preserving conformal isometry $\Psi:M\to M$. By
the hypothesis \eq{H} $\Psi$ can be extended to an isometry $\Phi$
of $b$ which leaves $\hyp$ invariant. Writing $\Upsilon$ as
$$\Upsilon = (\Upsilon\circ \Phi^{-1})\circ\Phi$$ we can decompose
$\Upsilon$ into an isometry of $b$ and a map
$\Upsilon\circ\Phi^{-1} $ which asymptotes to the identity. By
Corollary~\ref{C1} the metric $\Phi^*g$ satisfies the hypotheses
of Theorem~\ref{T1}, in the new coordinates $c_3$ as made precise
by that Corollary, so that the conclusions of Theorem~\ref{T1}
apply to $\Upsilon\circ \Phi^{-1}$. Let $b$ be the background
associated with the first coordinate system $c_1$, and let $\hb$
be that associated with $c_2$; since $\Phi$ is an isometry, the
background metric associated with $c_3$ coincides with that
associated with $c_1$. Now, by Theorem~\ref{TP1}, the integrals
\eq{toto} are invariant under the change of background which is
associated to $\Upsilon\circ\Phi^{-1}$:
\begin{equation}
  \label{eq:t1*}
 m(\hyp,g,b, X)= m(\hyp,g,\hb, \hat X)\;,\end{equation}
where the $\hb$-Killing vector $\hat X$ is associated to the
$b$-Killing vector field $X$ as described in the statement of
Theorem~\ref{TP1}. On the other hand, Lemma~\ref{L1} shows that
the
 isometry $\Phi$  reshuffles the integrals associated with
different Killing vectors,
\begin{equation}
  \label{eq:t1**}m(\hyp,\Phi^*g,b,(\Phi_*)^{-1}X) =
  m(\hyp,g,b,X)\;,\end{equation}
according to the action of $\Isot$ by push-forward on $\cK$, or
$\cKS$, or $\cKt$. (We note that since $\Phi$ is an $b$-isometry
preserving $\hyp$, $\Phi_*$ preserves the field of $b$-unit
normals to $\hyp$, hence the space $\cKS$ of those Killing vector
fields which are normal to $\hyp$. Similarly $\Phi$ preserves the
space $\cKt$ of Killing vector fields tangent to $\hyp$.)
Equations~\eq{eq:t1*}-\eq{eq:t1**} show that any invariant of the
action of $\Isot$ on $\cK$, or on $\cKS$,  or on $\cKt$, gives a
geometric invariant which can be associated to $\hyp$,
independently of the choice the coordinate
systems in $\cC$. 

When $b$ is the $(n+1)$-dimensional anti-de Sitter metric, the
relevant invariants based on Killing vector fields in $\cKS$ have
already been discussed in detail in the introduction,
Section~\ref{Si}. Consider the remaining Killing vector fields
$L_{(\mu)(\nu)}\in\cKt$, as given by \Eq{Lor2}. Equation~\eq{remK}
shows that under the action of $\Isot=O^\uparrow(1,n)$, the
orthochronous $(n+1)$-dimensional Lorentz group, the integrals
$$Q_{(\mu)(\nu)}\equiv m(\hyp,g,b,L_{(\mu)(\nu)})$$ transform as
the components of a two-covariant antisymmetric tensor. One  then
obtains a
 geometric invariant of $\hyp$ by calculating
\begin{equation}
  \label{eq:ch}Q\equiv Q_{(\mu)(\nu)}Q_{(\alpha)(\beta)}\eta^{(\mu)(\alpha)}\eta^{(\nu)(\beta)}\end{equation}
(for conventions see Appendix~\ref{Aai1}). In dimension $3+1$
another independent geometric invariant is obtained from
\begin{equation}
  \label{eq:ch1}
  Q^*\equiv
Q_{(\mu)(\nu)}Q_{(\alpha)(\beta)}\epsilon^{(\mu)(\alpha)(\nu)(\beta)}\;.
\end{equation}
In higher dimensions further invariants are obtained by
calculating $\text{tr}(P^{2k})$, $2\le 2k\le (n+1)$, where
$P^{(\alpha)}{}_{(\beta)}= \eta^{(\alpha)(\mu)}Q_{(\mu)(\beta)}$.
(In this notation $Q$ given by \Eq{eq:ch} equals
$\text{tr}(P^{2})$.) When $n+1$  is even one also has obvious
generalisations of \eq{eq:ch1}.

Consider, next, a (compact) strictly negatively curved $(\mn,h)$,
as considered in Appendix~\eq{Aai2}. In that case all Killing
vector fields are in $\cKS$, the action of $\Isot$ on $\cKS$ is
trivial, and all the geometric invariants of $\hyp$ given by
\eq{toto} are provided by the mass integrals considered in the
Introduction.

Let, finally, $(\mn,h)$ be a flat $(n-1)$-dimensional torus
$\mathbb{T}^n$, $n\ge 2$; as discussed in Appendix~\eq{Aai2}, all
conformal isometries  of $(\mn,h)$ are isometries and the action
of $\Isot$ on $\cK$ is trivial. It follows that in addition to the
mass we have  $n-1$ independent  invariants
$$m_A(\hyp,g)\equiv m(\hyp,g,b, \partial_A)$$
associated with  the Killing vectors $\partial_A$ of $(\mn,h)$;
here
 the $\partial_A$'s have been chosen so that they are
tangent to the $S^1$ factors of $\T^N=S^1\times\cdots\times S^1$,
and normalized to have unit length; such vector fields can loosely
be thought of as generating ``rotations'' of the torus
$\mathbb{T}^n$ into itself, giving the $m_A$'s an angular-momentum
type character.

\appendix
\section{The phase space and the Hamiltonians}
\label{AHam} {In \cite{ChAIHP} the starting
  point of the analysis is the Hilbert Lagrangian for vacuum Einstein
  gravity, $$\mathcal{L}= \sqrt{- \det g_{\mu\nu}}~\frac{g^{\alpha
      \beta} R_{\alpha \beta}}{16\pi}\;. $$ With our signature
  $(-+\cdots+)$ one needs to repeat the analysis in \cite{ChAIHP}
with
  $\mathcal{L}$ replaced by $$\frac{\sqrt{- \det
      g_{\mu\nu}}}{16\pi}\left(g^{\alpha \beta} R_{\alpha \beta}
    -2\Lambda\right)\;,$$ and without making the assumption $n+1=4$ done
  there; we follow the presentation in \cite{CJK}:
Consider the Ricci tensor, \be \label{JK1} R_{\mu\nu} =
\partial_\alpha \left[
  {\Gamma}^{\alpha}_{\mu\nu} - {\delta}^{\alpha}_{(\mu}
  {\Gamma}^{\kappa}_{\nu ) \kappa} \right] - \left[
  {\Gamma}^{\alpha}_{\sigma\mu} {\Gamma}^{\sigma}_{\alpha\nu}-
  {\Gamma}^{\alpha}_{\mu\nu} {\Gamma}^{\sigma}_{\alpha\sigma} \right]
\;, \ee where the $\Gamma$'s are the Christoffel symbols of $g$.
Contracting $R_{\mu\nu}$ with the contravariant density of metric,
\be \label{JK1.1}
\Kp^{\mu\nu} := 
\frac 1{16 \pi} \sqrt{-\det g } \ g^{\mu\nu} \;, \ee one obtains
the following expression for the Hilbert \Lagrangian density:
\begin{eqnarray}
{\tilde L} & = & \frac 1{16 \pi} \sqrt{-\det g } R = \Kp^{\mu\nu}
R_{\mu\nu} \nonumber \\ \label{JK3} & = &
\partial_\alpha \left[
\Kp^{\mu\nu} \left( {\Gamma}^{\alpha}_{\mu\nu} -
{\delta}^{\alpha}_{(\mu} {\Gamma}^{\kappa}_{\nu ) \kappa} \right)
\right] + \Kp^{\mu\nu} \left[ {\Gamma}^{\alpha}_{\sigma\mu}
{\Gamma}^{\sigma}_{\alpha\nu}-
 {\Gamma}^{\alpha}_{\mu\nu} {\Gamma}^{\sigma}_{\alpha\sigma} \right] \;.
\end{eqnarray}
Here we have used the metricity condition of $\Gamma$, which is
equivalent to the following identity: \be\label{JK2}
\Kp^{\mu\nu}_{\ \
  , \alpha} := \partial_\alpha \Kp^{\mu\nu} = \Kp^{\mu\nu}
{\Gamma}^{\sigma}_{\alpha\sigma} - \Kp^{\mu\sigma}
{\Gamma}^{\nu}_{\sigma\alpha} - \Kp^{\nu\sigma}
{\Gamma}^{\mu}_{\sigma\alpha} \;.  \ee Suppose now, that
${\Bgamma}^{\alpha}_{\sigma\mu}$ is another symmetric connection
in $M$, which will be used as a background (or reference)
connection. Denote by $\zR_{\mu\nu}$ its Ricci tensor.  From the
metricity condition \eq{JK2} we similarly obtain
\begin{eqnarray}
\Kp^{\mu\nu} \zR_{\mu\nu} \nonumber & = &
\partial_\alpha \left[
\Kp^{\mu\nu} \left( {\Bgamma}^{\alpha}_{\mu\nu} -
{\delta}^{\alpha}_{(\mu} {\Bgamma}^{\kappa}_{\nu ) \kappa} \right)
\right] - \Kp^{\mu\nu} \left[ {\Bgamma}^{\alpha}_{\sigma\mu}
{\Bgamma}^{\sigma}_{\alpha\nu}-
 {\Bgamma}^{\alpha}_{\mu\nu} {\Bgamma}^{\sigma}_{\alpha\sigma} \right]
 \nonumber \\
 &  & +
\Kp^{\mu\nu} \left[ {\Gamma}^{\alpha}_{\sigma\mu}
{\Bgamma}^{\sigma}_{\alpha\nu} + {\Bgamma}^{\alpha}_{\sigma\mu}
{\Gamma}^{\sigma}_{\alpha\nu} - {\Gamma}^{\alpha}_{\mu\nu}
{\Bgamma}^{\sigma}_{\alpha\sigma} - {\Bgamma}^{\alpha}_{\mu\nu}
{\Gamma}^{\sigma}_{\alpha\sigma} \right]
 \;. \label{JK4}
\end{eqnarray}
It is useful to introduce the tensor field \be \KA^\alpha_{\mu\nu}
:=\left( {\Bgamma}^{\alpha}_{\mu\nu} -
  {\delta}^{\alpha}_{(\mu} {\Bgamma}^{\kappa}_{\nu ) \kappa} \right) -\left({\Gamma}^{\alpha}_{\mu\nu} - {\delta}^{\alpha}_{(\mu}
{\Gamma}^{\kappa}_{\nu ) \kappa}\right) \label{Pend} \;.  \ee Once
the reference connection $\Bgamma^\alpha_{\mu\nu}$ is given, the
tensor $\KA^\alpha_{\mu\nu}$ encodes the entire information about
the connection $\Gamma^\alpha_{\mu\nu}$: \[
{\Gamma}^{\alpha}_{\mu\nu} = {\Bgamma}^{\alpha}_{\mu\nu} -
\KA^{\alpha}_{\mu\nu} + \frac 2{n} {\delta}^{\alpha}_{(\mu}
\KA^{\kappa}_{\nu ) \kappa}   \] (recall that the space-time
dimension is $n+1$). Subtracting Equation~\eq{JK4} from \eq{JK3},
and using the definition of $\KA^\alpha_{\mu\nu} $, we arrive at
the equation
\[\Kp^{\mu\nu}
R_{\mu\nu} -\frac{\sqrt{- \det
      g_{\mu\nu}}}{8\pi}\Lambda= -\partial_\alpha \left( \Kp^{\mu\nu}
\KA^\alpha_{\mu\nu} \right) + L \;,\] where
\begin{eqnarray*}  L &:=& \Kp^{\mu\nu}
\left[ \left(
    {\Gamma}^{\alpha}_{\sigma\mu} - {\Bgamma}^{\alpha}_{\sigma\mu}
  \right) \left( {\Gamma}^{\sigma}_{\alpha\nu}-
    {\Bgamma}^{\sigma}_{\alpha\nu} \right) - \left(
    {\Gamma}^{\alpha}_{\mu\nu} - {\Bgamma}^{\alpha}_{\mu\nu} \right)
  \left( {\Gamma}^{\sigma}_{\alpha\sigma} -
    {\Bgamma}^{\sigma}_{\alpha\sigma} \right) + \zR_{\mu\nu} \right]
\\ && -\frac{\sqrt{- \det
      g_{\mu\nu}}}{8\pi}\Lambda\;.\end{eqnarray*} This result may be
      used as follows: the quantity $L$ differs by a total divergence
      from the gravitational Lagrangian, and hence the associated
      variational principle leads to the same equations of motion.
      Further, the metricity condition \eq{JK2} enables us to rewrite
      $L$ in terms of the first derivatives of $\Kp^{\mu\nu}$: indeed,
      replacing in \eq{JK2} the partial derivatives $
\Kp^{\mu\nu}_{\ \;, \alpha} $ by the covariant derivatives
$\Kp^{\mu\nu}_{\ \ ; \alpha} $, calculated with respect to the
background connection $\Bgamma$,\be\label{JK2.1} \Kp^{\mu\nu}_{\ \
; \alpha} := \Kp^{\mu\nu} \left(
  {\Gamma}^{\sigma}_{\alpha\sigma} - {\Bgamma}^{\sigma}_{\alpha\sigma}
\right) - \Kp^{\mu\sigma} \left( {\Gamma}^{\nu}_{\sigma\alpha} -
  {\Bgamma}^{\nu}_{\sigma\alpha} \right) - \Kp^{\nu\sigma} \left(
  {\Gamma}^{\mu}_{\sigma\alpha} - {\Bgamma}^{\mu}_{\sigma\alpha}
\right) \;, \ee we may calculate $\KA^{\alpha}_{\mu\nu}$ in terms
of the latter derivatives. The final result is:
\begin{eqnarray}\label{Pend2}
  \KA^{\lambda}_{\mu\nu} & = &  \frac 12
  \Kp_{\mu\alpha}  \Kp^{\lambda\alpha}_{\ \ ; \nu}
  + \frac 12
  \Kp_{\nu\alpha}  \Kp^{\lambda\alpha}_{\ \ ; \mu}
  - \frac 12
  \Kp^{\lambda\alpha} \Kp_{\sigma\mu} \Kp_{\rho\nu}
  \Kp^{\sigma\rho}_{\ \ ; \alpha}
  \nonumber \\   & &
  +\frac 1{2(n-1)}
  \Kp^{\lambda\alpha} \Kp_{\mu\nu} \Kp_{\sigma\rho}
  \Kp^{\sigma\rho}_{\ \ ; \alpha} \;,
\end{eqnarray}
where by $\Kp_{\mu\nu}$ we denote the matrix inverse to
$\Kp^{\mu\nu}$; we assume that $n\ge 2$. Further,
\[ {\Gamma}^{\alpha}_{\mu\nu}-
 {\Bgamma}^{\alpha}_{\mu\nu} = - \KA^{\alpha}_{\mu\nu} + \frac 1{n-1}
 \Kp_{\sigma\rho}
  \Kp^{\sigma\rho}_{\ \ ; (\mu}{\delta}^{\alpha}_{\nu)} \;.  \]
We have \be\label{JK5alt} \frac{\partial L}{\partial
\Kp^{\mu\nu}_{\ \;, \lambda}} = \frac{\partial L}{\partial
\Kp^{\mu\nu}_{\ \ ; \lambda}}  = \frac{\partial L}{\partial
\Gamma^{\alpha}_{\beta\gamma}}\frac{\partial
\Gamma^{\alpha}_{\beta\gamma}}{\partial \Kp^{\mu\nu}_{\ \ ;
\lambda}} = \KA^{\lambda}_{\mu\nu}\;, \ee with the last equality
being obtained by tedious but otherwise straightforward algebra.
It follows that the tensor $ \KA^{\lambda}_{\mu\nu}$  is the
momentum canonically conjugate to the contravariant tensor density
$\Kp^{\mu\nu}$; prescribing this last object is of course
equivalent to prescribing the metric. Alternatively, one can
calculate
\begin{eqnarray}
L & =  & \frac 12 \Kp_{\mu\alpha} \Kp^{\mu\nu}_{\ \ ; \lambda}
\Kp^{\lambda\alpha}_{\ \ ; \nu} - \frac 14 \Kp^{\lambda\alpha}
\Kp_{\sigma\mu} \Kp_{\rho\nu} \Kp^{\mu\nu}_{\ \ ; \lambda}
\Kp^{\sigma\rho}_{\ \ ; \alpha} \nonumber \\ & & + \frac 18
\Kp^{\lambda\alpha} \Kp_{\mu\nu} \Kp^{\mu\nu}_{\ \ ; \lambda}
\Kp_{\sigma\rho}  \Kp^{\sigma\rho}_{\ \ ;
\alpha}+\Kp^{\mu\nu}\zR_{\mu\nu}  -\frac{\sqrt{- \det
      g_{\mu\nu}}}{8\pi}\Lambda\;.\label{defL}
\end{eqnarray}
and check  directly that \be\label{JK5}  \frac{\partial
L}{\partial \Kp^{\mu\nu}_{\ \ ; \lambda}} =
\KA^{\lambda}_{\mu\nu}\;. \ee

  Given a symmetric background connection
$B$ on $M$, we take $L$ given by Equation~\eq{defL} as the
\Lagrangian for the theory.  The canonical momentum $
\KA^{\lambda}_{\mu\nu}$ is defined by Equation~\eq{Pend2} or,
equivalently, by \Eq{JK5}. If ${\hyp}$ is any piecewise smooth
hypersurface in $M$, we define {\em  the space-time phase bundle
over $\hyp$ } as the collection of the $
(\KA^{\lambda}_{\mu\nu},\Kp^{\alpha\beta}) $'s over $\hyp$.  If $
(\delta_a\KA^{\lambda}_{\mu\nu},\delta_a \Kp^{\alpha\beta}) $,
$a=1,2$, are two sections over $\hyp$ of the bundle of vertical
vectors tangent to the space-time phase bundle, following
\cite{KijowskiGRG} we set \be\label{N.3g}
\Omega_{\hyp}((\delta_1\KA^{\lambda}_{\mu\nu},\delta_1
\Kp^{\alpha\beta}),(\delta_2\KA^{\lambda}_{\mu\nu},\delta_2
\Kp^{\alpha\beta}))=\int_{\hyp}
(\delta_1\KA^{\mu}_{\alpha\beta}\delta_2
\Kp^{\alpha\beta}-\delta_2\KA^{\mu}_{\alpha\beta}\delta_1
\Kp^{\alpha\beta})\rd S_\mu \, , \ee with the fields
$(\delta_1\KA^{\lambda}_{\mu\nu},\delta_1 \Kp^{\alpha\beta})$ and
$(\delta_2\KA^{\lambda}_{\mu\nu},\delta_2 \Kp^{\alpha\beta})$ such
that the integrals converge. Here $\rdS_{\mu}$ is defined as
\begin{equation}
  \label{eq:dSdef}
 \frac{\partial}{\partial x^\mu}\lrcorner \rd x^0 \wedge\cdots
\wedge\rd x^{n} \;,
\end{equation}
 where $\lrcorner$ denotes contraction.
This can be loosely thought of as being the ``symplectic'' form on
the gravitational phase space; however we will avoid this
terminology since the definition of a symplectic form involves
non-degeneracy conditions, which are quite subtle in an infinite
dimensional context, and which we do not want to address.

To be more specific, let $\hyp$ be a hypersurface which is the
union of a compact set with an asymptotic region $\hypext\approx
[R_0,\infty)\times M$ parameterized by $(r,v^A)$ as in the body of
this paper. Consider a background metric $b$ of the form
\eq{eq:a1} defined on $\hypext$, with its associated tetrad $e_a$;
we define the phase space $\cP_b$ as the space of those
smooth\footnote{The condition of smoothness of the relevant fields
is certainly not needed, and should be relaxed if an attempt is
made to obtain a full symplectic description of the situation at
hand.} sections $ (\KA^{\lambda}_{\mu\nu},\Kp^{\alpha\beta}) $
along $\hyp$ of the space-time phase bundle which satisfy the
following conditions:
\begin{enumerate}
\item[${\mycal C} 1$.]
First, we only allow those sections of the space-time phase bundle
which arise from solutions of the vacuum Einstein equations with
cosmological constant $\Lambda$ --- in particular the general
relativistic constraint equations with cosmological constant
$\Lambda$ have to be satisfied by the fields $
(\KA^{\lambda}_{\mu\nu},\Kp^{\alpha\beta}) $.
\item[${\mycal C} 2$.] Next, the $e_a$-tetrad
components of $g$ are required to be bounded on $\hypext$.
Moreover,  we impose the integral condition
\begin{equation}\int_\hypext  r\left(\sum_{a,b,c}
    |\znabla_ae^{bc}|^2 + \sum_{d,e} |e^{de}|^2\right) \; d\mu_b <
    \infty\;, \label{P1}\end{equation} where $e^{ab}$ are the
    $e_a$-tetrad components of $g-b$. In \eq{P1} $d\mu_b$ is, as
    before, the measure arising from the metric induced on $\hyp$ by
    the background metric $b$; in local coordinates such that
    $\hypext=\{t=0\}$ we have $d\mu_b=\sqrt{\det b_{ij}}drd^{n-1}v$,
    with the indices $i,j$ running from $1$ to $n$.
\item[${\mycal C} 3$.] Further,
the fall-off conditions
\begin{eqnarray}
 e^{ab}= o(r^{-n/2})\;, \quad e_c(e^{ab})= o(r^{-n/2})\;.
   \label{P1-}
\end{eqnarray}
 are assumed to hold on $\hypext$.
\item[${\mycal C} 4$.]Finally, we shall assume that the
following ``volume normalization condition'' is satisfied:
\begin{equation}\int_\hypext  r|b_{cd}e^{cd}| \; d\mu_b <
    \infty\;.  \label{P1+}\end{equation}
\end{enumerate}
(Recall that when $M$ is not parallelizable, then
    conditions \eq{P1}, \eq{P1-}, {\em etc.},  should be understood as
    the requirement of { existence of a covering of $M$ by a finite
    number of  open sets ${\mycal O}_i$ together
    with frames defined on $[R_0,\infty)\times {\mycal O}_i$
    satisfying the relevant conditions.})

Whenever we consider variations
$(\delta\KA^{\lambda}_{\mu\nu},\delta\Kp^{\alpha\beta}) $ of the
fields in $\cP_b$, we will require that those variations satisfy
the same decay conditions as the  fields in $\cP_b$.

From now on we shall assume that $B^\alpha_{\beta\gamma}$ is the
Levi-Civita connection of the background metric $b$. A condition
equivalent to \eq{P1}, and slightly more convenient to work with,
is
\begin{equation}\int_\hypext r \left(\sum_{a,b,c}
    |e_a(e^{bc})|^2 + \sum_{d,e} |e^{de}|^2\right) \; d\mu_b <
    \infty\;.  \label{P1equiv}\end{equation}
This follows immediately from \Eqsone{eq:b3} and \eq{eq:b4}, which
show that the $\znabla$-connection coefficients are bounded in the
frame $e_a$. It follows that  the fall-off conditions
\eq{C4}-\eq{C5} will ensure that ${\mycal C} 2$--${\mycal C} 4$
hold.

 Let us show that
\Eq{P1} guarantees that the integral defining $\Omega_\hyp$
converges. In order to see that, consider the identity:
\begin{eqnarray*}\znabla_{\alpha}e^{\mu\nu} & = &
\znabla_{\alpha}g^{\mu\nu} =
(\znabla_{\alpha}-\nabla_{\alpha})g^{\mu\nu}
\\ & = & -g^{\mu\sigma}(\Gamma^\nu_{\sigma\alpha} -
B^\nu_{\sigma\alpha})-g^{\nu\sigma}(\Gamma^\mu_{\sigma\alpha} -
B^\mu_{\sigma\alpha}) \;.
\end{eqnarray*}
The usual cyclic permutations calculation allows one to express
$\Gamma^\alpha_{\beta\gamma} - B^\alpha_{\beta\gamma}$ as a linear
combination of the $\znabla_{\alpha}e^{\mu\nu} $'s. It then
follows from \Eq{Pend} that the tetrad coefficients $p^a_{bc}$ of
$p^\alpha_{\beta\gamma}$ are, on $\hypext$, linear combinations
with bounded coefficients of the $\znabla_ae^{bc}$'s. In local
coordinates on $\hypext$ we have
$$\sqrt{|\det b_{\mu\nu}|}\sim r \sqrt{\det b_{ij}}\;,$$
hence
$$\int_{\hypext} |\delta p^t_{ab}|\;|\delta
\Kp^{ab}| dr d^{n-1}v \le C \sum_{a,b,c,d,e}\int_{\hypext} r
|\znabla_c\delta e^{de}|\;|\delta e^{ab}| d\mu_b <\infty\;.$$ Here
the coordinate $x^0\equiv t$ has been chosen so that
$\hypext=\{t=0\}$. Thus, $\Omega_\hyp$ is well defined on $\cP_b$,
as desired.

Recall, now, that $\Omega_\hyp$ coincides up to boundary terms
with the more familiar ``ADM symplectic form'' \cite{Kij1,Kij2}:
one sets
\begin{eqnarray} \label{Pkl}
  P^{kl} & := & \sqrt{\det g_{mn}} \ (\gthreeup^{ij}K_{ij}
  \gthreeup^{kl} - K^{kl} )\;,
\end{eqnarray}
where $K_{kl}$ is the extrinsic curvature of $\hyp$,
\begin{eqnarray} \label{Kl} K_{kl} & := & - \frac
  1{\sqrt{|g^{tt}|}} {\Gamma}^t_{kl}
\;,
\end{eqnarray}
with $\gthreeup^{kl}$ --- the three-dimensional inverse of the
induced metric $g_{kl}$ on $\hyp$; the indices on $K^{kl}$ have
been raised using $\gthreeup^{kl}$.   If we further choose the
coordinate $x^3$ in such a way that
$\partial\hypext=\{t=0,x^3=1\}$, then the ``symplectic'' form
(\ref{N.3g}) can be rewritten as~\cite{Kij1,Kij2}
\begin{eqnarray}\nonumber
\lefteqn{\Omega_{\hyp}((\delta_1 p^{\lambda}_{\mu\nu},\delta_1
\Kp^{\alpha\beta}),(\delta_2 p^{\lambda}_{\mu\nu},\delta_2
\Kp^{\alpha\beta})) =   \frac 1{16 \pi} \int_\hyp \left( \delta_1
g_{kl} \delta_2
  P^{kl} -\delta_2 g_{kl} \delta_1 P^{kl} \right) d^{n}x} \\ &
  \phantom{xxxxxxxxx} + \displaystyle\frac 1{16 \pi} \int_{\partial
  \hyp} \left(\delta_1 N^3 \delta_2 \frac{\sqrt{\det g_{kl}}}{N} -
  \delta_2 N^3 \delta_1 \frac{\sqrt{\det g_{kl}}}{N} \right)
  d^{n-1}v\;, \label{nN.3.1jj}\end{eqnarray} where
  $$N=1/\sqrt{-g^{tt}}\;,\quad N_k=g_{tk}\;,\quad
  N^3=(g^{3k}-\frac{g^{t3}g^{tk}}{g^{tt}})N_k\;.$$ Let us show that
  $\Omega_\hyp$ actually coincides with the ADM ``symplectic form'' on
  $\cP_b$. It clearly follows from \eq{P1} (with
  $e^{ab}$ replaced by $\delta e^{ab}$) that the volume integral there
  converges as before; it remains to show that the boundary integral
  vanishes.
We have
\begin{eqnarray*}
\delta \left( \frac {\sqrt{\det g_{ij}}} {N} \right) & = & \delta
\left(
    \sqrt{|\det g_{\mu\nu}|}\right)
\\
& = & \delta \left( \sqrt{\frac {\det g_{\mu\nu}}{\det
    b_{\mu\nu}}}\right) \sqrt{|{\det b_{\mu\nu}} |}
\\
&=& o(r^{-n/2})O(r^{n-1})= o(r^{n/2-1})\;.
\end{eqnarray*}
One easily checks the identity
$$ N^3= \frac {f_0(r)}{\sqrt{|g^{tt}|}} \;,$$
where $f_0$ is the future directed $g$-unit-normal to $\hyp$. We
have
\begin{eqnarray*}
  g^{tt}\equiv g(dt,dt) &=& (\eta^{ab}+e^{ab})e_a(t)e_b(t)
\\ &=& (\eta^{00}+e^{00})\left(e_0(t)\right)^2
\\
&=& (\eta^{00}+e^{00})|b^{tt}|\;,
\end{eqnarray*}
which gives
$$\frac 1 {\sqrt{|g^{tt}|}}=O(r)\;,$$
$$  \delta\left(
\frac 1 {\sqrt{|g^{tt}|}} \right) = \delta\left(
 \sqrt{\frac {b^{tt}}{g^{tt}}}\right)\frac 1{\sqrt{|b^{tt}|}} =
 o(r^{-n/2+1})\;.$$ Further, $$ f_0(r) = f_0{}^be_b(r) =
 f_0{}^1e_1(r)= o(r^{-n/2+1})\;,$$ $$\delta f_0(r) =\delta
 f_0{}^1e_1(r)= o(r^{-n/2+1})\;,$$ where $e_a$ is a $b$-orthonormal
 frame as in \Eq{frame}, and the vanishing of the boundary term in
 \eq{nN.3.1jj} readily follows.

According to \cite{KijowskiTulczyjew} (see also
\cite{CJK,KijowskiGRG}) the Hamiltonian associated with a one
parameter family of maps of the phase space into itself which
arise from  the flow of a vector field $X$ on the space-time
equals \be\label{N.5g} H(X,{\hyp}\przecinekJped )=\int_{\hyp}
(\KA^{\mu}_{\alpha\beta}\lie_X \Kp^{\alpha\beta} -X^\mu L)\rd
S_\mu \, \ee {\em provided that all the integrals involved are
well defined, and that the boundary integral in the variational
formula
\begin{eqnarray}\nonumber
  -\delta H
  & = & \int_{\hyp}\left(
    \cLX\KA^{\lambda}{_{\mu\nu}}\delta\Kp^{\mu\nu}-\cLX\Kp^{\mu\nu}
    \delta\KA^{\lambda}{_{\mu\nu}}\right) \rd S_\lambda
  \phantom{xxxxxxxxxxx} \\ & &  +
  \int_{\partial\hyp} X^{[\mu}{ \KA^{\nu ]}{_{\alpha\beta}} } \delta
  \Kp^{\alpha\beta} \rdS_{\mu\nu} \;,
\label{Hvar1g}
  \end{eqnarray}
vanishes.} In  the case when $\Bgamma$ is the metric connection of
a given background metric $\bmetric_{\mu\nu}$, and when $X$ is a
Killing vector field of $\bmetric_{\mu\nu}$, the identification
\begin{eqnarray}
  m(\hyp,g,b, X)&= &H(X,{\hyp}\przecinekJped )\;,
\label{toto2}\end{eqnarray} together with the calculations in
\cite{ChAIHP} leads to  \Eqs{toto}{Freud2.0}. More precisely, let
$E^{\nu\lambda}$ be given by the formula \cite{ChAIHP}
\begin{eqnarray}
  E^{\nu\lambda}&= &  \displaystyle{\frac{2|\det
  \bmetric_{\mu\nu}|}{ 16\pi\sqrt{|\det g_{\rho\sigma}|}}}
g_{\beta\gamma}(e^2 g^{\gamma[\nu}g^{\lambda]\kappa})_{;\kappa}
X^\beta \nonumber \\ && + \frac 1{8\pi} \sqrt{|\det
g_{\rho\sigma}|}~g^{\alpha[\nu}\delta^{\lambda]}_\beta
{X^\beta}_{;\alpha}
\;.\label{Freud2.01} 
\\
  \label{mas2.1}
e&=& {\sqrt{|\det
g_{\rho\sigma}|}}/{\sqrt{|\det\bmetric_{\mu\nu}|}}\; .
\end{eqnarray}
It can be checked that all the formulae of
\cite[Appendix~B]{ChAIHP} are dimension independent, and lead to
the identity \be\label{N.5g1} E^\lambda :=
\KA^{\lambda}_{\alpha\beta}\lie_X \Kp^{\alpha\beta} -X^\lambda L =
E^{\nu\lambda}{}_{;\nu} +T^\lambda{}_\kappa X^\kappa\;, \ee where
the matter energy-momentum tensor has been defined in \eq{N.5g2}.
Now, when $b$ is the anti de Sitter metric, the integral of
$E^\lambda dS_\lambda$ over large ``balls'' $B_R:=\{r\le R\}$
within $\hyp$ would diverge if we tried to pass with the radius of
those balls to infinity because we have $$ E^\lambda\Big|_{g=b}=
-(\zRs-2\Lambda)X^\lambda/16\pi\;,$$ with $\zRs$
--- the Ricci scalar of the background metric $b$, and
$\zRs-2\Lambda=4 \Lambda/(n-1)$ in an $(n+1)$-dimensional
space-time. We therefore add to $E^\lambda$ a $g$-independent term
which will cancel this divergence: indeed, such terms can be
freely added to the Hamiltonian because they do not affect the
variational formula that defines a Hamiltonian.  {}From an energy
point of view such an addition corresponds to a choice of the zero
point of the energy. We thus set $$ \ourU^{\alpha\beta}:=
E^{\alpha\beta}-E^{\alpha\beta}\Big|_{g=b} \;.$$ From the
definition of $E^\lambda$ and from \Eq{N.5g1} one easily finds
\begin{eqnarray}
\nonumber
  16\pi\znabla_\beta \ourU^{\alpha\beta} &= &    \left( \sqrt{|\det g|}
    g^{ab} - \sqrt{|\det b|}    b^{ab}\right) \zRm_{ab} X^\beta
\\ & &  +2 \Lambda \left(\sqrt{|\det b| }- \sqrt{|\det g| }\right)X^\beta
\nonumber \\ & & +
\left(\mathring{T}^\lambda{}_\kappa-T^\lambda{}_\kappa\right)X^\kappa
\nonumber \\ & & + \sqrt{|\det    b|} \;\left( Q^\alpha{}_{\beta}
X^\beta + Q^\alpha{}_{\beta \gamma}
    \znabla^\beta X^\gamma\right)\;, \label{C3}\end{eqnarray} where
    $Q^\alpha{}_{\beta}$ is a quadratic form in $e_a(e^{bc})$, and
    $Q^\alpha{}_{\beta \gamma}$ is bilinear in $e_a(e^{bc})$ and
    $e^{ab}$, both with bounded coefficients. Further,
$\mathring{T}^\lambda{}_\kappa$ is defined as in \Eq{N.5g2} with
$g$ replaced by $b$.

From now on we assume that both $g$ and $b$ are Einstein, and we
only consider vector fields $X$ which are $b$-Killing vector
fields and satisfy \be \label{A1.}|X|+ |\zn X| \le C r \ee
 for some constant $C$; this holds for
all the backgrounds considered in Appendix~\ref{Aiso}, in
particular for the generalized Kottler metrics \eq{Kottler2}.
Theorem~\ref{T0} then shows that the integral defining $H$
converges for fields in $\cP_b$.

Suppose, further, that the $b$-Killing vector field $X$ has the
property that {\em the associated variations of the fields are
compatible with the boundary conditions} imposed on fields in
$\cP_b$. This means in particular that we must have
\begin{equation}\int_\hyp  r\sum_{a,b,c}
    | \lie_X \left(\znabla_a e^{bc}\right)|^2 \;
d\mu_b < \infty\;.
    \label{eqP2}\end{equation}
  Clearly the volume
    integral in the variational formula \eq{Hvar1g} converges under
\eq{eqP2} together with the remaining conditions set forth above.
Further,  the boundary integral there vanishes under \eq{P1-}, so
that \Eq{N.5g} does indeed provide the required Hamiltonian on
$\cP_b$.

For Killing vectors satisfying \eq{A1.} \Eq{eqP2} will hold if
\begin{equation}\int_\hyp  r^3\sum_{a,b,c,d}
    | \left(\znabla_a \znabla_d e^{bc}\right)|^2 \; d\mu_b < \infty\;,
    \label{eqP2.1}\end{equation} but we emphasize that the weaker
    condition \eq{eqP2} suffices.

\section{Isometries and Killing vectors of the background}\label{Aiso}

\subsection{$(n+1)$-dimensional anti-de Sitter metrics}
\label{Aai1}

\newcommand{\Y}{{\mycal Y}}

For $n\ge 2$ consider the $(n+1)$-dimensional anti-de Sitter
space-time $(\cM,b)$, thus $b$ is given by \eq{eq:a1a}
with $h$ --- the unit round metric on the $(n-1)$-dimensional
sphere ${}^{(n-1)}S$. As elsewhere we set $\hyp = \{t=0\}.$ When
$\mn$ is the two-dimensional sphere, the Killing vectors of $b$
are given in \cite{HT}. For higher dimensional spheres the
$b$-Killing vector fields are easily found by thinking of $b$ as
the metric induced on the covering space of the hyperboloid
\begin{equation}
  \label{hyp}
  \eta_{(a)(b)} y^{(a)}y^{(b)} =-\ell^2
\end{equation}
in the $(n+2)$-dimensional manifold $\Y$ with the
metric\footnote{$\Y$ can be identified with the universal covering
of the space obtained by removing the set $y^{(0)}=y^{(n+1)}=0$
from $\R^{n+2}$; $\Y$ then inherits the local coordinates
$y^{(a)}$ used in \Eq{locy}. However, in order to understand the
geometry of $\cM$ in a neighbourhood of $\hyp$ it is sufficient
--- and most convenient --- to  think of $\Y$ as of $\R^{n+2}$.}
\be\label{locy} \eta_{(a)(b)}dy^{(a)}dy^{(b)} = -(dy^{(0)})^2 +
\sum_{(i)=(1)}^{(n)}(dy^{(i)})^2 -(dy^{(n+1)})^2\;. \ee Throughout
this section the indices $(a),(b)$, \emph{etc.,\/} run from $ (0)$
to $(n+1)$. The hyperboloid can be locally parameterized by
coordinates $t$, $x^i$ implicitly defined by the
equations\footnote{The spherical coordinates associated to the
  ``cartesian'' coordinates $x^i$ give the form (\ref{eq:a1a}) of the
  metric $b$.}
\begin{equation}
\label{transf1} y^{(0)} = \ell\cos (t/\ell)\sqrt{1+r^2/\ell^2} \;,
\end{equation}
\begin{equation}
\label{transf2} y^{(n+1)} = \ell \sin (t/\ell)\sqrt{1+r^2/\ell^2}
\;,
\end{equation}
\begin{equation}
\label{transf3} y^{(i)} =x^i\;,
\end{equation}
with $r^2 = \sum_{i=1}^n (x^i)^2$, where $x^i=r n^i$, and
$n^i\in{}^{(n-1)}S$ can eventually be expressed in terms of
coordinates on the sphere ${}^{(n-1)}S$. For example, for $n=3$ we
can use $x^1=r \sin(\theta)\cos(\varphi)$, $x^2 =r \sin(\theta)
\sin(\varphi)$, $x^3= r \cos(\theta)$, with $\theta,\varphi$ ---
the usual spherical coordinates.  It is also convenient to
represent the hypersurface $\hyp \subset \cM$ given by $\hyp
=\{t=0\}$ as $\{\eta_{(a)(b)} y^{(a)}y^{(b)} =-\ell^2\}\cap
\{y^{(n+1)} =0, y^{(0)}
>0\} \subset \Y$. We set
\begin{equation}
  \label{Lor2}
L_{(a)(b)} = y_{(a)} \frac{\partial~~}{\partial y^{(b)}} -y_{(b)}
\frac{\partial~~}{\partial y^{(a)}}\;,
\end{equation}
where $y_{(a)}=\eta_{(a)(b)} y^{(b)}$. The $L_{(a)(b)}$'s are
Killing vector fields of $(\Y,\eta_{(a)(b)})$. Further they are
tangent to the hyperboloid $\{\eta_{(a)(b)} y^{(a)}y^{(b)}
=-\ell^2\}$ and hence define, by restriction, Killing vector
fields of the hyperboloid with the induced metric. In fact they
span the space of all the Killing vectors of $b$ because there is
the right maximal number of them.  From the coordinate
transformation (\ref{transf1})-(\ref{transf3}) one can compute the
corresponding Killing vectors of anti de Sitter space-time in the
coordinates $\{t,x^i\}$, obtaining
$$
L_{(n+1)(0)} = \ell\frac{\partial }{\partial t}\;,
$$
$$
L_{(i)(n+1)} = \frac{x^i}{\sqrt{1+r^2/\ell^2}} \cos(t/\ell)
\frac{\partial }{\partial t} + \ell\sqrt{1+r^2/\ell^2}
\sin(t/\ell)
\frac{\partial}{\partial x^i}\;,
$$
$$
L_{(i)(0)} = -\frac{x^i}{\sqrt{1+r^2/\ell^2}} \sin(t/\ell)
\frac{\partial }{\partial t} +\ell\sqrt{1+r^2/\ell^2} \cos(t/\ell)
\frac{\partial}{\partial x^i}\;,
$$
$$
L_{(i)(j)} = x^i \frac{\partial}{\partial x^j} - x^j
\frac{\partial}{\partial x^i}\;.
$$
Let $\cKS$ be the set of Killing vector fields of $b$ which are
orthogonal to $\hyp$; from the expressions above it is not
difficult to check that a vector basis for $\cKS$ is given by
$\frg_{(\mu)} = L_{(n+1)(\mu)}|_{t=0}$, where $(\mu)$ runs from
$(0)$ to $ (n)$.

\begin{Proposition}
\label{app-lemma1} Let $\Phi:\cM \to \cM$ be an isometry of $b$
such that $\Phi(\hyp)=\hyp$.  Then there exists  a Lorentz
transformation matrix $\Lambda^{(\nu)}{}_{(\mu)}$ such that the
basis vectors of $\cKS$ satisfy
$$\Phi_{*}\frg_{(\mu)} = \Lambda^{(\nu)}{}_{(\mu)} \frg_{(\nu)}\;.$$
\end{Proposition}
\remark We note that the property $\Phi_*(\cKS) =\cKS$ follows
from the fact that $\Phi$ preserves $\hyp$, which implies that
$\Phi$ maps the field of unit normals to $\hyp$ into itself.

\medskip

\proof As is well known, for every isometry $\Phi:\cM \to \cM$ of
$b$ there exists a diffeomorphism $\hat{\Phi}:\Y \to \Y$, isometry
of $\eta_{(a)(b)}$, such that $\Phi$ is the restriction of
$\hat{\Phi}$ to the hyperboloid \eq{hyp}. In  coordinates we have
\begin{equation}
  \label{Lor0}\hat{\Phi}^{(a)}(y) =
  \Lambda^{(a)}{}_{(b)}y^{(b)}\;,\end{equation}
where $\Lambda^{(a)}{}_{(b)}$ is a matrix satisfying
\begin{equation}
  \label{Lor}
  \Lambda^{(c)}{}_{(a)} \Lambda^{(d)}{}_{(b)} \eta_{(c)(d)} =
\eta_{(a)(b)}\;.
\end{equation}
The hypersurface $\hyp \subset \Y$ is given by
$\eta_{(a)(b)}y^{(a)}y^{(b)} =-1$ and $y^{(n+1)} =0$ together with
the condition $y^{(0)} >0$, so that the condition
$\hat{\Phi}(\hyp) =\hyp$ implies
$$
\Lambda^{(a)}{}_{(b)} = \left[
\begin{array}{cc}
\Lambda^{(\mu)}{}_{(\nu)} & 0 \\
0 & \pm 1
\end{array}
\right]\;,
$$
where we split the indices as $(a)=(\mu), (n+1)$.
Equation~\eq{Lor} shows that $\Lambda^{(\mu)}{}_{(\nu)}$ is a
$n+1$-dimensional Lorentz transformation,
$(\Lambda^{(\mu)}{}_{(\nu)})\in O(1,n)$. Equations~\eq{Lor0} and
\eq{Lor2} imply that under push-forward by $\hat{\Phi}$ the
Killing vectors of $\eta_{(a)(b)}$ transform as
$$
\hat{\Phi}_{*}L_{(a)(b)} = \Lambda^{(c)}{}_{(a)}
\Lambda^{(d)}{}_{(b)} L_{(c)(d)}
$$
in particular, the basis vectors of $\cKS$ transform as
\begin{eqnarray*}
\hat{\Phi}_{*}\frg_{(\mu)} &=&
\hat{\Phi}_{*}L_{(n+1)(\mu)}\Big |_{\hat{\Phi}(y^{(n+1)}) =0} \\
&=& \Lambda^{(c)}{}_{(n+1)} \Lambda^{(d)}{}_{(\mu)}
L_{(c)(d)}\Big |_{y^{(n+1)} =0} \\
&=& \pm \Lambda^{(\nu)}{}_{(\mu)} \frg_{(\nu)}.
\end{eqnarray*}
Replacing $\Lambda^{(\nu)}{}_{(\mu)}$ by
$-\Lambda^{(\nu)}{}_{(\mu)}$ if necessary, the result follows.
 \qed

 Let $\cKt$ be the space of $b$- Killing vectors spanned by the
 $L_{(\mu)(\nu)}$'s, thus $\cKt$ contains all the $L_{(a)(b)}$'s
 which are not in $\cKS$.  An identical calculation as in the proof
 above shows that under isometries of $b$ preserving $\hyp$ we have
\begin{eqnarray} \label{remK}
\hat{\Phi}_{*}L_{(\mu)(\nu)} &=&
 \Lambda^{(\sigma)}{}_{\mu} \Lambda^{(\rho)}{}_{(\nu)}
L_{(\sigma)(\rho)}\;.
\end{eqnarray}
It follows 
that the resulting representation of the Lorentz group on $\cKt$
is equivalent to a representation on two-contravariant
anti-symmetric tensors.


\subsection{$h$'s with a non-positive Ricci tensor} \label{Aai2}
We consider metrics \eq{A1}, as in Section~\ref{S2}. In what
follows we shall only consider $(M,h)$'s with a non-positive Ricci
curvature, with $n\ge 3$, the case $n=2$ being covered by the
previous section. We shall further assume that the scalar
curvature $R_h$ of $h$ (the Ricci scalar) is a constant. We note
that the vector fields
\begin{equation}
  \label{A5}
  X=X^0 n= {\lambda\over a}n = \lambda \partial_t \;,\qquad \lambda \in \R\;,
\end{equation}
where $n=e_0$ is the field of future pointing unit normals to the
hypersurfaces $\{t=\mathit{const}\}$, are  Killing vector fields
for the metric $b$ whatever $a=a(r)$. The non-vanishing connection
coefficients, $$\zo^{a}{}_{bc}\equiv \theta^a(\zn_{e_c}e_b)\;,$$
with respect to this frame  are
\begin{equation}
  \label{A2}
  \zo^A{}_{1B}=  \frac 1 {ra(r)} \delta ^A_B= -\zo^1{}_{AB}\;,\quad
  \zo^A{}_{BC}= \frac 1r \beta^A{}_{BC}\;, \quad \zo_{100}=-\zo_{010}
  = -{a'(r)\over a^2(r)}\;,
\end{equation}
where the $\beta^A{}_{BC}$ are the
 Levi-Civita connection coefficients of $h$ with respect to  the frame
 $\alpha^A$.
 The $AB$ components of the Killing equations read
\begin{equation}
  \label{eq:A1}
  \caD_A X_B + \caD_B X_A + {1\over  a(r)}h_{AB}X^1 =0\;,
\end{equation}
where $\caD$ is the covariant derivative operator associated with
the metric $h$, which shows that $X^B\partial_B$ is a conformal
Killing vector field on $M$. Uniqueness of solutions of the
volume-normalized Yamabe equation in the case under consideration
implies that conformal Killing vector fields of $(\mn,h)$ are
necessarily Killing vectors, hence
$$X^1\equiv 0\;.$$
The $00$, $01$ and $0A$ components of the Killing equations read
\begin{eqnarray}
  \label{eq:A2}
  e_0(X_0)& = & 0\;,
\\  \label{eq:A3}
  e_1(X_0)+ {a'\over a^2} X_0& = & 0\;,
\\  \label{eq:A4}
  e_A(X_0)+e_0(X_A)& = & 0\;.
\end{eqnarray}
Equation~\eq{eq:A2} shows that $X_0$ is $t$-independent.

Suppose, first, that the Ricci tensor of $h$ is strictly negative.
It is well known\footnote{\label{Kf}The Killing equations imply
  $X_B\caD_C\caD^C X^B =
  -R_{AB} X^AX^B$, where $R_{AB}$ is the Ricci tensor of $h$;
integration of this equation over $\mn$ shows that $X^A$ is
covariantly constant when $R_{AB}$ is non-positive, and vanishes
when $R_{AB}$ is strictly negative.} that in this case $(\mn,h)$
has no non-trivial Killing vector fields so that $X^A\equiv 0$,
and Equation~\eq{eq:A4} shows that $X_0$ is $v^A$-independent.
Integrating \eq{eq:A3} yields then the  one parameter family of
Killing vector fields \eq{A5}, which shows that the algebra of all
Killing vector fields of $b$ is one-dimensional.

Suppose, next, that $(M,h)$ is an $(n-1)$-dimensional flat torus
$(\T^{n-1},\delta)$. Then the $X^A$'s are covariantly
constant$^{\mathrm{\ref{Kf}}}$ vector fields on $\T^{n-1}$, which
shows that $X^A=X^A(t,r)$ in coordinates $v^A$ in which the metric
$\delta$ has constant entries. Integrating \eq{eq:A4} over
$\T^{n-1}$ gives
$$0= \int_{\T^{n-1}} e_A(X_0) = -e_0(X_A)\;\mathrm{Vol}(\T^{n-1})\;,$$
hence $X_A=X_A(r)$. Equation~\eq{eq:A4} implies then that $X_0$ is
$v^A$-independent, so that $X_0=X_0(r)$,  from \eq{eq:A3} we
recover \eq{A5}, and $\cKS$ is again one-dimensional, as claimed.
We note that the $1A$ component of the Killing equation implies
that the $X^A$'s are in fact $r$-independent, which gives a
complete description of the set of Killing vector fields occurring
in this case.

The above arguments extend to all manifolds with constant scalar
curvature and non-positive Ricci curvature, as follows: suppose
that $(\mn,h)$ has non-trivial Killing vector fields. The $1A$
component of the Killing equations gives
$$e_1(X_A) - {1\over ra} X_A = 0 \quad \Longrightarrow \quad X^A =
X^A(t,v^A)\;.$$
 Integration of Equation~\eq{eq:A3} gives
 \begin{equation}
   \label{A6}
   X^0 = {\lambda(v^A)\over a(r)}\;,
 \end{equation}
for some function $\lambda$ on $\mn$. Equation~\eq{A5} inserted
into \eq{eq:A4} gives
$$e_A(\lambda)=-a^2(r)\partial_t X^A\;,$$
which is compatible with \Eq{A5} only if $\partial_t X^A=0$, hence
$e_A(\lambda)=0$. Summarizing, we have proved

\begin{Proposition}
  \label{PA1} If $(\mn,h)$ has non-positive Ricci curvature and
  constant scalar curvature, then all Killing vector fields of the
  metric $b$ given by \Eq{A1} are of the form
$$X={\lambda \over a} n + X^A(v^B)\partial _A\;, \qquad\lambda \in
\R\;,$$ where $X^A(v^B)\partial _A$ is a Killing vector field of
the metric $h$.
\end{Proposition}

\section{Equality of the Hamiltonian mass with the Abbott-Deser one}
\label{AAD}

In this appendix we consider a subset  $\R\times \Sigma_{\ext}$ of
a four dimensional space-time $(\cM,g)$ defined by
 a coordinate system $\{x^{\alpha}\}$; we identify
$\Sigma_{\ext}$ with the set $ \{x^{\alpha} : x^0 =0\}$. The space
 coordinates $(x^i)$ on $\Sigma_{\ext}$ will be written as $(r,v^A)$,
 with the range of $r$ being $[R_0,\infty)$, and with the $v^A$ being
 local coordinates on some compact two dimensional manifold. Assume that
there exists a frame $\{ e_a\}_{a=0}^3$ defined on
$\Sigma_{\ext}$, which defines a background metric
$b^{\alpha\beta} = \eta^{ab} e_a^{\alpha}e_b^{\beta}$, where
$\eta^{ab} = \mbox{diag}(-1,1,1,1)$. In other words, the tetrad
$\{e_a\}$ is orthonormal with respect to $b_{\alpha\beta}$. Assume
that in this frame the space-time metric $g$ has the form
$$g_{ab} = \eta_{ab} + e_{ab}\;,$$
where $$e_{ab} = o(1/r^{\alpha})\:,\qquad e_a(e_{bc}) =
o(1/r^{\alpha}) 
\;,$$ for some $\alpha>0$.  The Abbott-Deser mass $M_{AB}$
associated with $X$ is defined as \cite{AbbottDeser}
\begin{equation}
\label{m-ab} M_{AB} =  \frac{1}{2} \lim_{R \to \infty}
\int_{\partial \Sigma_R} V^{\alpha\beta} dS_{\alpha\beta}\;,
\end{equation}
where
\begin{equation}
\label{m-ab1} V^{\alpha\beta}(h)= \frac{1}{8\pi}\, b \left(
K^{\alpha\beta\sigma\kappa}{}_{;\kappa} X_{\sigma} -
K^{\alpha\kappa\sigma\beta} X_{\sigma;\kappa} \right) \;, \ee with
$$K^{\alpha\beta\sigma\kappa} =
b^{\alpha[\kappa}H^{\sigma]\beta} +
H^{\alpha[\kappa}b^{\sigma]\beta}\;,\quad H^{\alpha\beta} =
e^{\alpha\beta} -\frac{1}{2} e_{\gamma}{}^{\gamma}
b^{\alpha\beta}\;,$$
$$ b= \sqrt{|\det b_{\mu\nu}|}\;.$$
 Note that
$\tK^{\alpha\beta\sigma\kappa}$ has the same symmetries as the
Riemann tensor. Let $\ourU^{\alpha\beta}$  be the ``Hamiltonian
superpotential'' defined by  \eq{Fsup2new}; assume that
\begin{equation}
\label{m-ab2} (|X| + |\nabla X|)b = O(r^\beta)\;, \ee
 we then
claim that
$$\ourU^{\alpha\beta} = \myvareps
\,\tV^{\alpha\beta} + \oepsb\;.$$ In order to establish this,
recall that
\begin{eqnarray*}
\det(g_{\alpha\beta}) &=&
\det\left(b_{\alpha\gamma}\left[\delta^{\gamma}{}_{\beta} +
\myvareps
b^{\gamma\sigma}\th_{\sigma\beta} + \oeps\right]\right) \\
&=& \det(b_{\alpha\gamma}) \det(\delta^{\gamma}{}_{\beta} +
\myvareps e^{\gamma}{}_{\beta})\;,
\end{eqnarray*}
where $e^{\gamma}{}_{\beta} = b^{\gamma\sigma}\th_{\sigma\beta} +
\oeps$. A well known identity gives
$$
\det(g_{\alpha\beta}) = \det(b_{\alpha\beta}) \left(1 + \myvareps
\, \th_{\gamma}{}^{\gamma} + \oeps \right)\;.
$$
Let us write $\ourU^{\alpha\beta} =\ourU^{\alpha\beta}{}_{\gamma}
X^{\gamma} + \widehat{\ourU}^{\alpha\beta}$, where
$$
\ourU^{\alpha\beta}{}_{\gamma} := \frac{1}{8\pi} \frac{b}{e}
\left(e^2 g^{\sigma[\alpha} g^{\beta]\kappa}\right)_{;\kappa}
g_{\gamma\sigma}\;,
$$
$$
\widehat{\ourU}^{\alpha\beta} := \frac{1}{8\pi} \left( \sqrt{|\det
g_{\rho\sigma}|} \; g^{\kappa[\alpha}\delta^{\beta]}{}_{\gamma} -
b \; b^{\kappa[\alpha} \delta^{\beta]}{}_{\gamma} \right)
X^{\gamma}{}_{;\kappa} \;.
$$
We have
$$
e^2 = 1+ \myvareps \,\th_{\alpha}{}^{\alpha} + \oeps \;,$$ so that
the first term above can be written as
\begin{eqnarray*}
{\ourU}^{\alpha\beta}{}_{\gamma} &=& \frac{1}{8\pi} b \left[ (1+
\myvareps \th_{\alpha}{}^{\alpha} + \oeps) \left(b^{\sigma[\alpha}
b^{\beta]\kappa} - \myvareps \,
b^{\sigma[\alpha}\th^{\beta]\kappa} - \myvareps
\,\th^{\sigma[\alpha} b^{\beta]\kappa}
+ \oeps\right)\right]_{;\kappa} b_{\gamma\sigma}\\
&=& \frac{b}{8\pi} \,\myvareps \, b_{\gamma\sigma} \left(
\th_{\rho}{}^{\rho} \,b^{\sigma[\alpha} b^{\beta]\kappa} -
b^{\sigma[\alpha} \th^{\beta]\kappa} - \th^{\sigma[\alpha}
b^{\beta]\kappa} \right)_{;\kappa}
+ \oepsb \\
&=&- \frac{b}{8\pi}\, \myvareps \,b_{\gamma\sigma} \left(
b^{\sigma[\alpha} \tH^{\beta]\kappa} + \tH^{\sigma[\alpha}
b^{\beta]\kappa} \right)_{;\kappa} + \oepsb
\end{eqnarray*}
so that
$$
{\ourU}^{\alpha\beta}{}_{\gamma} = \frac{b}{8\pi} \, \myvareps\,
b_{\gamma\sigma} \tK^{\alpha\beta\sigma\kappa}{}_{;\kappa} +
\oepsb\;.
$$
Similarly,
\begin{eqnarray*}
\widehat{\ourU}^{\alpha\beta} &:=& \frac{1}{8\pi} \left(
\sqrt{|\det g_{\rho\sigma}|} \;
g^{\kappa[\alpha}\delta^{\beta]}{}_{\gamma} - b \;
b^{\kappa[\alpha} \delta^{\beta]}{}_{\gamma} \right)
X^{\gamma}{}_{;\kappa}\\
&=& \frac{1}{8\pi} \left( \sqrt{|\det g_{\rho\sigma}|} \;
g^{\kappa[\alpha}b^{\beta]\gamma} - b \; b^{\kappa[\alpha}
b^{\beta]\gamma} \right)
X_{\gamma;\kappa} \\
&=& \frac{b}{8\pi} \left[ \left(1+ \frac{1}{2}\, \myvareps \th +
\oeps\right) \left(b^{\kappa[\alpha}b^{\beta]\gamma} - \myvareps
\, \th^{\kappa[\alpha}b^{\beta]\gamma} + \oeps\right)
- b^{\kappa[\alpha}b^{\beta]\gamma} \right] X_{\gamma;\kappa}\\
&=& - \frac{b}{8\pi} \,\myvareps
\tH^{\kappa[\alpha}b^{\beta]\gamma} X_{\gamma;\kappa}  + \oepsb\;.
\end{eqnarray*}
Now, $X_{\gamma}$ is a Killing vector of $b_{\mu\nu}$, therefore
\begin{eqnarray*}
\tH^{\kappa[\beta}b^{\alpha]\gamma} X_{\gamma;\kappa}
&=& \tH^{\kappa[\beta}b^{\alpha]\gamma}X_{[\gamma;\kappa]}\\
&=& \frac{1}{2} \left(\tH^{\kappa[\beta}b^{\alpha]\gamma}
- \tH^{\gamma[\beta}b^{\alpha]\kappa} \right) X_{\gamma;\kappa}\\
&=& \frac{1}{2} \tK^{\kappa\gamma\alpha\beta}X_{\gamma;\kappa}\;,
\end{eqnarray*}
so, we have obtained
$$
\widehat{\ourU}^{\alpha\beta} = \frac{b}{16\pi} \, \myvareps
\tK^{\kappa\gamma\alpha\beta}X_{\gamma;\kappa} + \oepsb\;.
$$
The two terms together give
\begin{eqnarray*}
\ourU^{\alpha\beta}&=& \frac{b}{8\pi}\, \myvareps  \left(
\tK^{\alpha\beta\sigma\kappa}{}_{;\kappa} X_{\sigma} +
\frac{1}{2}\,\tK^{\kappa\gamma\alpha\beta}X_{\gamma;\kappa}\right)
+ \oepsb\\
&=& \frac{b}{8\pi}\, \myvareps  \left(
\tK^{\alpha\beta\sigma\kappa}{}_{;\kappa} X_{\sigma} -
\frac{1}{2}\,\tK^{\alpha\beta\gamma\kappa}X_{\gamma;\kappa}\right)
+ \oepsb\;.
\end{eqnarray*}
As $\tK^{\alpha\beta\gamma\kappa}$ has the same symmetries as the
Riemann tensor, we have $\tK^{\alpha[\beta\gamma\kappa]} =0$,
which implies that $\frac{1}{2} \, \tK^{\alpha\beta\gamma\kappa} =
\tK^{\alpha[\kappa\gamma]\beta}$, and
\begin{eqnarray*}
\ourU^{\alpha\beta}&=& \frac{b}{8\pi}\, \myvareps  \left(
\tK^{\alpha\beta\sigma\kappa}{}_{;\kappa} X_{\sigma} -
\tK^{\alpha[\kappa\gamma]\beta}X_{\gamma;\kappa}\right)
+ \oepsb\\
&=& \myvareps \, \tV^{\alpha\beta} + \oepsb\;.
\end{eqnarray*}
If \be\label{abcond} \beta - 2 \alpha \le 0\;,\ee we obtain
equality of the Abbott-Deser mass with the Hamiltonian one; recall
that $\beta=n$ for the anti-de Sitter type metrics considered in
the body of the paper, and \Eq{abcond} reproduces the condition
$\alpha \ge n/2$,  identical to that which arises in the proof of
coordinate-invariance of the mass integral.

Summarizing, we have proved:

\begin{Proposition}
Suppose that \Eqsone{m-ab1}, \eq{m-ab2} and \eq{abcond} hold. Then
the Hamiltonian mass coincides with the Abbott-Deser one; in
particular, either they are both undefined, or both diverge, or
both converge to the same values.
\end{Proposition}

\textbf{Acknowledgements:} PTC acknowledges useful discussions
with S.~Ilias, A.~Polombo and A.El Soufi. We appreciated the
friendly hospitality of the Max Planck Institute for Gravitation
in Golm during final stages of work on this paper.


\addcontentsline{toc}{1}{\bigskip \noindent  {\bf References \hfill}}

\bibliographystyle{amsplain}
\bibliography{
../../references/newbiblio,%
../../references/reffile,%
../../references/bibl,%
../../references/Energy,%
../../references/hip_bib,%
../../references/netbiblio}
\end{document}